\documentclass[pra,showpacs,preprintnumbers,amsmath,amssymb,tightenlines,twocolumn,superscriptaddress]{revtex4}

\usepackage{graphicx}
\usepackage{dcolumn}
\usepackage{bm}
\usepackage{epsfig}
\usepackage{stmaryrd}
\usepackage{color}

\newcommand{\bra}[1]{\langle\,{#1}\, |}
\newcommand{\ket}[1]{|\,{#1}\,\rangle}

\newcommand{\vek}[1]{\boldsymbol{#1}}
\newcommand{\rref}[1]{Ref.~\cite{#1}}

\newcommand{\eref}[1]{Eq.~(\ref{#1})}

\newcommand{\cref}[1]{chapter~\ref{#1}}

\newcommand{\Cref}[1]{Chapter~\ref{#1}}

\usepackage{ulem}  
\begin{document}

\title{Quantum scattering with semiclassical asymptotic motion}

\author{John S.\ Briggs}
\affiliation{Institute of Physics, University of Freiburg, Freiburg, Germany and \\
Department of Physics, Royal University of Phnom Penh, Cambodia}
\email{briggs@physik.uni-freiburg.de}

\author{James M.\ Feagin}
\affiliation{Department of Physics, California State University-Fullerton, Fullerton, CA 92834, USA }
\email{jfeagin@fullerton.edu}

\begin{abstract}
Conventional scattering theory is incomplete in that it does not adequately describe the behaviour of the wave function at
macroscopic distances from the scattering reaction volume. In scattering experiments particles are incident from sources at macroscopic distance and measured at macroscopic distance from the microscopic reaction volume.  Standard {\emph{experimental}} procedure is to describe asymptotic particle motion by classical mechanics. However, this transition to classical mechanics is not accounted for in standard quantum scattering theory. Hence, we revisit conventional  scattering theory with the introduction of semiclassical propagators to describe the wave function in both incident and outgoing asymptotic scattering channels. This description leads to the manifestation of classical trajectories in the quantum wave function both before and after the collision. It includes the case of extraction with external fields as
in modern multi-particle detectors. We derive the asymptotic Kemble \emph{imaging theorem} (IT) limit to demonstrate how formal quantum scattering theory contains implicit classical mechanics threaded throughout the asymptotic wave functions.
\end{abstract}

\maketitle

\section {Introduction}
Quantum scattering theory addresses one of the fundamental problems of quantum mechanics, namely, the description of the collision of microscopic particles. This involves their initial production
in sources at macroscopic distance from their
region of interaction and their subsequent motion from this  reaction volume of, at most, atomic or molecular dimensions to a detector positioned usually at macroscopic distance
from the reaction volume. Standard scattering theory and text-book expositions of it, have changed little since its inception in the 1950s \cite{LS,GMandG}. Time-independent  scattering theory applies only to time-independent potentials of interaction. In its simplest form, two-body potential scattering, one begins with the time-independent Schr\"odinger equation (TISE) {\emph{at fixed energy}} and defines, from the asymptotic form of the spatial Green function leading to a corresponding asymptotic form of the spatial wave function, a scattering amplitude (or equivalently a transition matrix element)
for transition to a final momentum state. Assuming constant incident flux of particles this is converted into a cross section

It is interesting that the standard general scattering theory, as presented in Goldberger and Watson \cite{GW}, for example, begins not with a fixed energy in the TISE but with the time-dependent Schr\"odinger equation (TDSE). However, the time is then eliminated by taking, in a somewhat involved procedure, the limits at infinite times before and after the collision. This restores the fixed energy such that the resulting elements of the theory, for example the transition matrix, the scattering matrix and M\o ller operators, are all time independent and expressed in terms of the time-independent interaction potentials. Recently \cite{BriggsFeaginGer}, following Gerjuoy \cite{Gerjuoy}, we have given a derivation of time-independent scattering theory which avoids the redundant introduction and elimination of time. This theory follows the text-book derivation of time-independent potential scattering but generalized to multi-particle, multi-channel collisions.

We point out three features of modern experiments involving fragmentation and re-arrangement collisions that render the standard scattering theory inadequate for the interpretation of the results of these scattering processes:

(i) In connection with ``atto- and femto-second" physics, modern experiments use light sources providing extremely short, intense pulses. Since this involves the approximation of the light source by a classical field, the potential of interaction is explicitly time-dependent. Furthermore, the results depend upon the pulse details, and the time-independent theory with the associated definition of a cross section based on steady beams cannot be used.

(ii) In the standard time-independent theory asymptotically free motion of colliding particles at large distances from the microscopic collision region is assumed. However, modern experimental procedures use external fields to extract and manipulate charged particles emitted from the collision volume and free asymptotic motion does not apply.

(iii) To interpret the flight of particles to and from macroscopic distances with respect to the extent of the microscopic collision region, experimenters routinely assume that classical mechanics is valid, even for light particles such as electrons. This classical element is nowhere to be found in time-independent scattering theory, which remains fully quantum mechanical.

 The use of classical mechanics automatically introduces time variables into the dynamical description and, for many particles, different times of detection apply generally. 
Despite this use of classical mechanics, the measurement of particles in coincidence exhibits clear quantum properties of coherence, interference, 
and entanglement thus signalling the preservation of the quantum wave function out to macroscopic distances. One aim of our formulation of scattering theory is to justify how these apparently conflicting aspects of experiments can be reconciled.

Our aim in this paper is to remedy the above three deficiencies of fixed-energy theory, the latter two of which are deficient also in the time-dependent theory.
We suggest that a formulation of scattering theory, both for time-dependent and time-independent interactions, based upon direct time propagation with appropriate Hamiltonians is both simpler
and easier to connect with modern experimental methods.

 The collision automatically separates into three time zones, before, during and after the interaction region. The novelty is that we use semi-classical 
approximations for the wave functions before and after the collision and the full quantum propagator only during the microscopic interaction region.  In this formulation, the key quantity describing the result of the collision is not the transition matrix, as in standard theory, but rather the  collision-complex momentum-space wave function.

 A time-dependent version of  scattering theory does exist of course, for example see \cite{Roman}, and in various forms has been used particularly where incident beams of photons or heavy ions are approximated as providing {\emph{classical}} electric or magnetic fields. However, here also a transition matrix approach is used most often. We advocate a direct time propagation, as used increasingly in modern numerical calculations, showing the importance of the momentum-space wavefunction emanating from the collision region, as has been emphasised by 
 Macek \cite{Macek}.
 
  Clearly, for time-varying interactions only the time-dependent propagation is applicable. We suggest that also for time-independent potentials it provides a simple and direct formulation, avoiding some of the problems, for example, of energy-dependent Green functions, infinite time limits and scattering wave functions extending over all space, inherent to the time-dependent theory.
 
  We  begin by showing that the generalisation of a result known as the ``imaging theorem"  (IT) allows a different formulation of scattering theory in the asymptotic zone, both before and after the collision. The central result of this imaging theorem, first proved in 1937 by Kemble \cite{Kemble}  to give a classical meaning to the momentum variable, is a semiclassical description of the wave function  at asymptotically large distance. Indeed, the imaging theorem involves precisely a description of the passage of particles from the reaction volume to a distant detector. According to the theorem the quantum wave function is preserved to macroscopic distance but its variables, position or momentum, are connected by classical mechanics. This immediately resolves the dichotomy of point (iii) above.

As we will show, the IT gives a new perspective to the quantum description of scattering. After propagating over macroscopic distances, either before or after collision, the wave function variables describe allowed classical trajectories and semiclassical quantum mechanics is valid. Only within the microscopic reaction volume, where forces vary over atomic-scale distances, is a full quantum description required.
The momentum-space wave function of the collision complex emerging from the reaction volume emerges naturally from the IT.

The imaging theorem was so-named since it shows that the asymptotic spatial wave function is the image of the momentum wave function right at the edge of the collision volume. Although perhaps a little presumptuous to accord this result the status of a theorem, nevertheless we will retain the nomenclature and refer to the result described below as the IT.

The IT has been proven and used in scattering theory several times since (and largely independent of) Kemble's derivation. Exclusively it is used to describe the asymptotic state of {\emph{ free }} motion of particles following collision.
Examples are in electron impact excitation \cite{ITfree} or in ion-atom collisions \cite{Groz} to propagate a calculated wave function to large times \cite{Macek}. Other derivations are in describing interfering electron trajectories \cite{Kleber} or in connection with Bohmian mechanics \cite{Holland, Allori}.
The IT also lies at the basis of Dollard's ``scattering into cones" theorem \cite{Dollard, Newton}, again applicable to the free motion of particles following the collision. We generalise this result to include motion in extracting fields.

We concentrate firstly on the time development of particle motion after the quantum collision process has occurred and the outgoing particle momenta are already fixed. The collision reaction occurs in a microscopic volume about the origin $\vek r = 0$ and particles are detected after typically macroscopic times (e.g.\ nanoseconds). \emph{We take as $t=0$ the moment the quantum collision begins} and $t=\tau$ when it has ended and all reaction fragments have exited the microscopic reaction volume. This collision duration is nevertheless of the order of atomic times ($\sim \!10^{-16} \, \mbox{s}$). 

\begin{figure}[t]
\includegraphics[scale=.45]{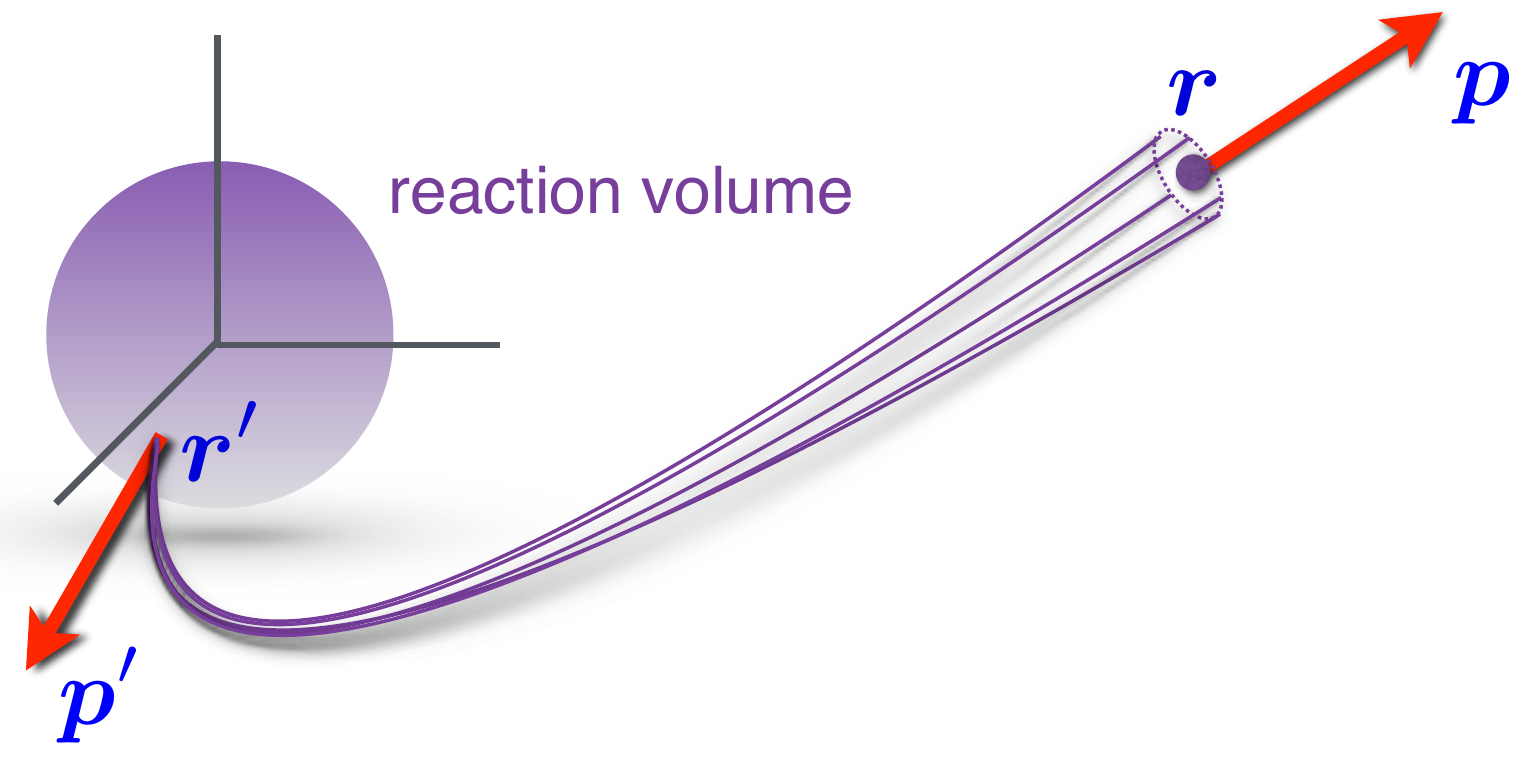}
\caption{\label{fig1} Extraction of a particle exiting a scattering reaction into a bundle of possible classical trajectories originating at $\vek r'$ with initial momentum near $\vek p'$ and ending near a distant detector at the point $\vek r$ with momentum $\vek p$. }
\end{figure}

As derived originally, the IT is overtly time dependent. It relates the asymptotic spatial wave function $\Psi(\vek r,t)$, where the coordinates $\vek r$ specify the macroscopic detection positions of all $N$ emanating particles at the same  time $t$, to the momentum wave function $\tilde\Psi(\vek p', \tau)$ of all $N$ particles at the outer edge of the microscopic reaction volume near $\vek r \approx 0$ at time $\tau \ll t$. Its essence
illustrated in Fig.\ \ref{fig1} is most simply stated in the {\emph{quantum}} result
\begin{equation}
\label{ITquantt}
|\Psi(\vek r, t)|^2 \approx  \frac{d\vek p'}{d\vek r}\,  |\tilde\Psi(\vek p', \tau)|^2, 
\end{equation}
where $d\vek p'/d\vek r$ however is the {\emph{classical}} density of outgoing trajectories.
Thus, the crucial feature of the IT is that the position and momentum variables in the quantum wave function are not independent but are connected by classical mechanics. That is, each initial momentum  of particle $n$ in the set $\vek p'$ at time $\tau$ determines the asymptotic positions $\vek r$ at large final time $t \gg \tau$. 

This result is to be contrasted with the asymptotic form derived from time-independent theory, say for the simple case of potential scattering of two particles. There one has
 \begin{equation}
 \label{TIasy}
|\Psi(\vek r,E)|^2 \approx \frac{1}{r^2} |f(\vek p',E)|^2,
\end{equation}
where $f(\vek p',E)$ is the amplitude for scattering into a state of fixed momentum $\vek p'$ and fixed energy $E$ .
In energy representation, the variables $\vek r$ and $\vek p$ are fully quantum and one simply defines the directional correspondence
$\vek p' \equiv p' \hat{\vek r}$ . Later in the paper we will connect the two forms of \eref{ITquantt} and \eref{TIasy}.

The key to the derivation of the IT lies in {\emph{semiclassical}} quantum dynamics. We have shown \cite{FeaginBriggs_ITII} that this allows a generalisation of the IT to external-field extraction, thereby resolving point (ii) above. More importantly, it shows how the use of classical motion for individual particles can be reconciled with the manifestations of quantum 
coherence, interference, and entanglement at large distances from the reaction volume.
A result of this paper is to derive, for time-independent interactions, a time-independent IT and to expose the implicit asymptotic classical aspects of time-independent theory, which are not usually made apparent in text books. 

The approach to the semiclassical and classical limits of quantum theory has a long history stretching back to the inception of quantum mechanics \cite{VV, Wheeler} and can be traced through many early applications in nuclear, atomic and molecular scattering.
The general theory of semiclassical quantum mechanics, also for bound states, was assembled more recently \cite{Gutz,Berry,Schulman,GottfriedNEW}. The specific application to the semiclassical theory of collisions is given by Miller \cite{Miller}, by Pechukas \cite{Pech1}, and by Rost and Heller  \cite{RostHeller,Rost}, for example.

However, these theories are concerned with the limiting treatment of the {\emph{complete scattering process}}, connecting ultimately to a classical collision cross section. As such the classical limit appears when the collision energy is large compared to all potential energies of interaction operating in the problem under consideration. This large energy guarantees that the action (\emph{energy $\times$ time}) or (\emph{momentum $\times$ position}) has a value much greater than $\hslash$ {\emph{at all times}.}
 
 It is important to distinguish this from the limit considered here. We treat the collision complex as wholly quantum and only the asymptotic motion in the semiclassical approximation. Then the IT becomes valid, at all collision energies, when the system wave function has been propagated to {\emph{distances and times}} such that the action appearing in the wave function has a value much in excess of $\hslash$.

The structure of the paper is as follows. 
In section II the IT in its usual time form is presented with emphasis on the semiclassical nature of the wave function in the asymptotic zone and includes the case of external field extraction of particles emanating from the collision volume.
This theory is suited particularly to ``attosecond" physics involving experiments with short intense laser pulses. 

In section III we treat a time-dependent collision process by employing the IT also to describe the {\emph{incident channel} wave function.
In section IV we develop a new derivation of time-independent scattering theory, again using the semiclassical approximation of the IT in both final and initial scattering channels. Although our derivation here also includes external field extraction, we connect with standard text-book results to show that a cross section of the usual form can be derived from our semiclassical asymptotic description.
 
Section V contains commentary on the implications of the IT semiclassical wave function for the interpretation of the results of coincidence experiments in terms of entanglement, coherence and interference arising from many-particle interactions.

Since the contrast of time and energy representations of scattering theory is a main topic of this paper, we give in the Appendix A a brief overview of the time and energy treatments of both quantum and classical mechanics. We compare the time-dependent and timeless versions of the equations of motion and the connection of classical and quantum mechanics by the semiclassical limit in both cases.

For simplicity of notation, in most of the paper, the theory is presented for the laser dissociation into two final particles or to the collision of two structureless particles, leading to the same two particles after the collision. The extension of the scattering theory to describe the dissociation or collision of composite particles, involving in general particle re-arrangement and multiple fragments in the final channel is given in Appendix B. The time-dependent theory, extended to include particles detected at different arrival times, is in section B1 and, for completeness, the time-independent general theory is described in section B2.

\section{Time-dependent scattering theory with a semiclassical asymptote}

\subsection{ Semiclassical propagation and the Kemble IT limit}

Kemble's aim in introducing the IT was not so much to prove the validity of free classical asymptotic motion as to give ``an operational definition of momentum for free particles."
He points out that a momentum measurement on a single particle requires in principle two measurements of position and time. One registers initially $\vek r', t'$ and later
position $\vek r$ at time $t$. This gives the velocity
\begin{equation}
\vek v' = \frac{\vek r - \vek r'}{t - t'}.
\end{equation}
Nevertheless he argues that if $\vek r'$ is confined to microscopic variation, then defining the velocity  simply as $\vek v' = \vek r /t$ ``in which $r$ is the measured position at some time $t$ great enough so that $r$ is large compared to the uncertainty in its measurement" will allow momentum to be measured with essentially arbitrary accuracy. 
That is, Kemble takes effectively microscopic $\vek r' = 0$ and microscopic $t' = 0$.  Then he uses a stationary phase argument to justify identifying the momentum calculated from this classical velocity with that appearing in the momentum-space quantum wave function, that is, ``a relation between measured momenta and the wave numbers of Fourier analysis."

This operational definition of momentum via classical velocity has been assumed implicitly ever since to identify measured momenta with those defined in standard scattering theory.

The IT has been employed historically in scattering theory mostly to extract information from the asymptotic post-collision form of numerically-calculated wave functions propagated in time \cite{Macek}.
 This emphasis on the final state is particularly appropriate to strong, short laser pulses interacting with atoms and molecules. In typical experiments a particle beam is crossed with a laser beam so that the photon flux and particle beam density characterise the initial conditions. The experimenter then measures the number of particles impinging on the detector as a function of position and time of flight. 

 Taking as time $t = 0$ the moment the collision begins and time $t = \tau$ when all particles have exited the reaction volume after the collision, then the
subsequent propagation of the quantum state follows the equation
\begin{equation}
\label{propfinal}
\ket{\Psi(t)} =  U_F(t,\tau) \,\ket{\Psi(\tau)}, 
\end{equation}
where $U_F(t,\tau)$ is the time-development operator defined outside the reaction volume and the subscript $F$ denotes the possibility of  using extraction fields.
Projecting onto an eigenstate $\bra{\vek r}$ of final position and inserting a complete set of momentum eigenstates $\ket{\vek p}$, this propagation is expressed in terms of wave functions as
\begin{equation}
\label{Psiprop}
\Psi(\vek r,t)  = \int  d\vek p \,\tilde K_F(\vek r,t; \vek p,\tau) \,  \tilde\Psi(\vek p,\tau), 
\end{equation}
where $\tilde K_F(\vek r,t; \vek p,\tau) = \bra{\vek r}U_F(t,\tau)\ket{\vek p}$ is the mixed coordinate-momentum propagator. 

In semiclassical approximation this mixed propagator can be expressed in the form
 \begin{equation}
 \label{Kpcl}  
 \tilde K_F(\vek r,t; \vek p,\tau) =  \frac{1}{(2\pi\hslash)^{3/2}}\Big|{\rm{det}} \frac{\partial^2\tilde S}{\partial \vek r\partial\vek p}\Big|^{1/2} ~e^{i \tilde S(\vek r,t;\vek p,\tau)/\hslash},
\end{equation}
where $\tilde S(\vek r,t;\vek p,\tau)$ is the {\emph{classical}} action of \eref{TDcl}. The tilde serves to denote that here we use the $(\vek r, \vek p)$ representation.
(In the case of a single outgoing particle considered here, $\partial^2\tilde S/\partial \vek r\partial\vek p$ is a $3 \times 3$ matrix).

For simplicity, we suppress a sum over possible multiple trajectories connecting $\vek r'$ with $\vek r$ in the fixed time $t$ but for different energies. Additionally, we suppress a Maslov index that counts the number of conjugate points where individual trajectories intersect one another along caustics \cite{Gutz, Berry, Schulman}. 

The integral over all quantum $\vek p$  in \eref{Psiprop} is evaluated for $t \gg \tau$ in the stationary phase approximation (SPA). The stationary phase point occurs at the 
point $\vek p = \vek p'$ where the condition
\begin{equation}
\label{SPAp}
\frac{ \partial \tilde S}{\partial \vek p} = 0 
\end{equation}
is satisfied. Note that this corresponds to initial $\vek r' \approx 0$  in line with Kemble's argument. The result of the integral in SPA is (for details see \rref{BriggsFeaginNJP})
\begin{equation}
\label{IT_mom}   
\Psi(\vek r, t)  \approx (i)^{-3/2} \left(\frac{d\vek p'}{d\vek r}\right)^{1/2} ~e^{i  S(\vek r,t;0,\tau)/\hslash} \, \tilde\Psi(\vek p',\tau),
\end{equation}
where $S(\vek r,t;\vek r',\tau)$ is the classical action in position space, connected to $\tilde S(\vek r,t;\vek p',\tau)$ by the Legendre transformation $S(\vek r,t;\vek r', \tau) = \tilde S(\vek r,t;\vek p, \tau) - \vek p \cdot \vek r'$,
which also clarifies \eref{SPAp}.  

To appreciate the implications of this IT result, it is important to illustrate the result of the two approximations

 a) semiclassical propagator and 
 
 b) SPA  evaluation of the momentum integral. 
 
In the exact full quantum expression \eref{Psiprop} the variables $\vek r$ and $\vek p$ are independent and the wave functions in the two spaces connected only by the integral kernel,
essentially a generalised Fourier transformation. Hence, in principle, each and every possible $\vek p$ value, initial momentum, corresponds to a given $\vek r$ value of final position.

Approximation of the kernel by its semiclassical form connects the two coordinates via classical mechanics. For example, for free motion and $t \gg \tau$, one has $\vek r \approx  \vek r' +\vek p'\,t/m $. Thus, if measurement fixes 
final position $\vek r$ and time of flight $t$, one has still the integral over varying $\vek p'$ values or equivalently varying initial position $\vek r' $. 

The SPA approximation removes the integral and
fixes the $\vek r'$ to a single value, in this case to zero. This isolates a single fixed momentum $\vek p' \approx m \vek r/t$ for each measured $\vek r$ and time $t$. Thus, the crucial feature of the IT is that the position and momentum variables in the {\emph{quantum}} wave function are not independent but are connected by {\emph{classical}} mechanics. That is, each initial momentum of particle $n$ in the set $\vek p'$ at time $\tau $ determines the asymptotic position $\vek r$ at large final time $t \gg \tau$.
This is the essential content of \eref{IT_mom}. 

The IT result relies upon the action greatly exceeding $\hbar$.  In \rref{BriggsFeaginNJP} we showed that this occurs after distances that  are large compared to atomic dimensions but 
 still microscopic compared to the size of  typical experimental apparatus. Thus any quantum system propagating to distances, and correspondingly over times, that are macroscopic undergoes an autonomous transition to a state described by a semiclassical wave function. 
 
 It should  be made clear that for the particular cases of free motion or TI extracting fields with corresponding potentials at most quadratic in the coordinates, both the semiclassical approximation and the SPA evaluation of the momentum integral are {\emph{exact}}. This perhaps explains why the semiclassical nature of the asymptotic motion has been overlooked until now. Also, this justifies many classical or semiclassical treatments of the motion of particles in a variety of experiments. A prime example is the use of classical trajectories to describe atomic motion in experiments of the Stern-Gerlach type. 
Another example is the continuum motion of photo-electrons in a strong laser field of attoseconds duration. Here electrons initially moving out from the residual ion can be turned around in the oscillating field and return to re-scatter from the parent ion leading to high-harmonic generation \cite{Kul}. The IT justifies the approximate use of classical mechanics to describe such trajectories.

From \eref{IT_mom},
the IT can be written in the form
   \begin{equation}
  \label{locus1}
|\Psi(\vek r,  t)|^2 \approx  \frac{d\vek p'}{d\vek r} \, |\tilde\Psi(\vek p',\tau)|^2.
\end{equation}
Note that the asymptotic quantum probability density $|\Psi(\vek r,  t)|^2$ appears as a product of two factors. One is the quantum probability density $|\tilde\Psi(\vek p', \tau)|^2$ of initial momentum $\vek p'$ resulting from the collision. The other factor factor is the {\emph{classical}} density of trajectories defined by the Jacobian determinant
\begin{equation}
\label{VVdet}
 \left| \det \frac{\partial\vek p'}{\partial\vek r} \right| = \frac{d\vek p'}{d\vek r}, 
\end{equation}
called the van Vleck factor \cite{VV}. Its importance in semiclassical mechanics has been emphasized particularly by Gutzwiller \cite{Gutz}, who referred to this density of trajectories factor as $C(t)$, see also Berry and Mount \cite{Berry}.
Outside the reaction volume, a particle emanating from a volume $d\vek p'$ around $\vek p'$ will arrive in a volume $d\vek r$ around $\vek r$.
One sees, remarkably, that quantum propagation outside the reaction volume is described by a purely classical quantity. 

 One can also write \eref{locus1} as
  \begin{equation}
  \label{locus2}
|\Psi(\vek r,  t)|^2\,d\vek r \approx |\tilde\Psi(\vek p',\tau)|^2\, d\vek p'.
\end{equation}
This shows that the locus of points of equal detection probability lie along classical trajectories. 
This result is also an elemental version of Dollard's ``scattering into cones theorem" \cite{Dollard}.

Standard time-independent scattering theory provides generally cross sections differential in fixed outgoing particle momenta. 
These cross sections can be compared to measurements only by defining constant particle fluxes calculated from time-independent wave functions. Although in  \rref{BriggsFeaginGer} and in the preceding section, the emphasis is on particle detection at fixed positions, the theory is seemingly at variance with modern coincident detection of
differing particles at different times. Indeed the timing of each particle is necessary to define the classical momenta. 

A single global time is usually used to connect the time-independent and time-dependent theories via Fourier transform of the time propagator for example. In appendix A we present a modified scattering theory for non-interacting particles in the outgoing channel. In this theory, following the IT, each particle propagates in its own time along an individual  classical trajectory to the detector.

\subsection{The Kemble IT limit from a Taylor expansion in spatial coordinates}

The restriction to $\vek r' \approx 0$ means that the  IT can be derived also using the propagator in the full coordinate representation. Then the asymptotic wave function appears as an integral over all initial quantum variables $\vek r'$,
\begin{equation}
\label{Psiprop2}
\Psi(\vek r,t)  = \int  K_F(\vek r,t; \vek r',\tau) \, \Psi(\vek r',\tau) \, d\vek r',
\end{equation}
Again, the propagator can be represented in the semiclassical form \cite{Gutz}
\begin{equation}
\label{Krr'}
\begin{split}
 K_F(\vek r,t; \vek r',\tau) & =  \frac{1}{(2\pi i\hslash)^{3/2}}\Big|{\rm{det}} \frac{\partial^2 S}{\partial \vek r\partial\vek r'}\Big|^{1/2} ~e^{i  S(\vek r,t;\vek r',\tau)/\hslash}\\
 & =  \frac{1}{(2\pi i\hslash)^{3/2}}\left(\frac{d\vek p'}{d\vek r}\right)^{1/2}~e^{i  S(\vek r,t;\vek r',\tau)/\hslash},
 \end{split}
\end{equation} 
where, as before, $S(\vek r,t;\vek r',\tau)$ is the classical action in coordinate space. 

Now it is recognised that for times $t \gg \tau $ the $\vek r'$ integral is confined to a small volume around $\vek r' \approx 0$, so that the action can be expanded around this point as
\begin{equation}
\label{Taylor}
S(\vek r,t;\vek r',\tau) \approx  S(\vek r,t;0,\tau)  +  \frac{\partial S}{\partial \vek r'}\Big|_0\cdot \vek r'.
\end{equation}
Then, with $ \partial S/\partial \vek r'|_0  \equiv -\vek p'$, substitution in the integral \eref{Psiprop2} gives a Fourier transform and the result 
\begin{equation}
\label{IT_coor}    
\Psi(\vek r, t)  \approx (i)^{-3/2} \left(\frac{d\vek p'}{d\vek r}\right)^{1/2} ~e^{i S(\vek r,t;0,\tau) /\hslash} \, \tilde\Psi(\vek p', \tau),
\end{equation}
which is the IT of \eref{IT_mom}.

Thus we have two, essentially conjugate, derivations of the IT. The first recognises the finite spread in momentum space of the scattering wave function at $t =  \tau$. Then the SPA in the
momentum integral of \eref{Psiprop} leads to well-defined classical momenta for $t = \tau$} in the scattering wave function at the exit of the collision region. 
Correspondingly, the position-space scattering wave function for $t \approx \tau$ has essentially zero extent so that the $\vek r'$ integral of \eref{Psiprop2} can be approximated around $\vek r' \approx 0$ to give the same momentum wave function as in the IT result of  \eref{IT_mom}.

\subsection{Asymptotic Free Motion}

The IT derived by Kemble \cite{Kemble} and others \cite{ITfree,Groz,Macek,Kleber, Holland,Allori} is for asymptotically free motion. 
For free motion, the mixed propagator of \eref{Kpcl} is simply
 \begin{equation}
 \label{freepropcl}
 \tilde K(\vek r,t; \vek p,\tau) = \frac{1}{(2\pi \hslash)^{3/2}}~e^{i \tilde S(\vek r,t;\vek p,\tau)/\hslash},
\end{equation}
where  $\tilde S(\vek r,t;\vek p,\tau) = \vek p \cdot \vek r - p^2 T/(2m)$ is the classical action for free motion with $T = t - \tau$.

Hence, $ \tilde K(\vek r,t; \vek p,\tau)$ is a free-particle momentum wave function and \eref{Psiprop} for free motion becomes just the  Fourier transform between coordinate and momentum wave functions.
Here the SPA condition \eref{SPAp} gives exactly the classical free-motion condition
\begin{equation}
\frac{ \partial \tilde S}{\partial \vek p}\Big|_{\vek p'} \equiv \vek r' = \vek r - \vek p' T/m = 0
\end{equation}
so that $\vek p' = m (\vek r - \vek r')/T $ the free-particle classical momentum.

The coordinate space propagator for free motion is
\begin{equation}
\label{Kposition}
\begin{split}
 K(\vek r,t; \vek r', \tau)& = \frac{m}{(2\pi i\hslash T)^{3/2}}~e^{i S(\vek r,t;\vek r', \tau)/\hslash}\\
& = \frac{m}{(2\pi i\hslash T)^{3/2}}\exp{\left[i\frac{m}{2\hslash T} (\vek r - \vek r')^2   \right]},
\end{split}
\end{equation}
since $S(\vek r,t;\vek r', \tau) = m (\vek r - \vek r')^2 /(2T)$ 
is the classical action in coordinate space.
This result is obtained from \eref{freepropcl} with the Legendre transformation 
$S(\vek r,t;\vek r', \tau) = \tilde S(\vek r,t; \vek p', \tau) - \vek p' \cdot \vek r'$ 
along with $\vek p' = m (\vek r - \vek r')/T$. 

One has as required $\vek p' \equiv -\partial S/\partial \vek r'$
 giving the van Vleck factor
\begin{equation}
\label{freeVVdet}
\frac{d\vek p'}{d\vek r} = \left| \det \frac{\partial\vek p'}{\partial\vek r}  \right|  = \left(\frac{m}{T} \right)^3.
\end{equation}
Then, for free motion in the limit $t \gg \tau,  T \approx t $, one has from \eref{IT_mom} or \eref{IT_coor}  the IT
 \begin{equation}
 \label{kembleIT}
\Psi(\vek r, t)  \approx \left(\frac{m}{i t}\right)^{3/2} e^{ i m r^2/(2\hslash t)} \, \tilde\Psi(\vek p',\tau),
\end{equation}
 as derived originally by Kemble.
 
\begin{figure}[b]
\includegraphics[scale=.35]{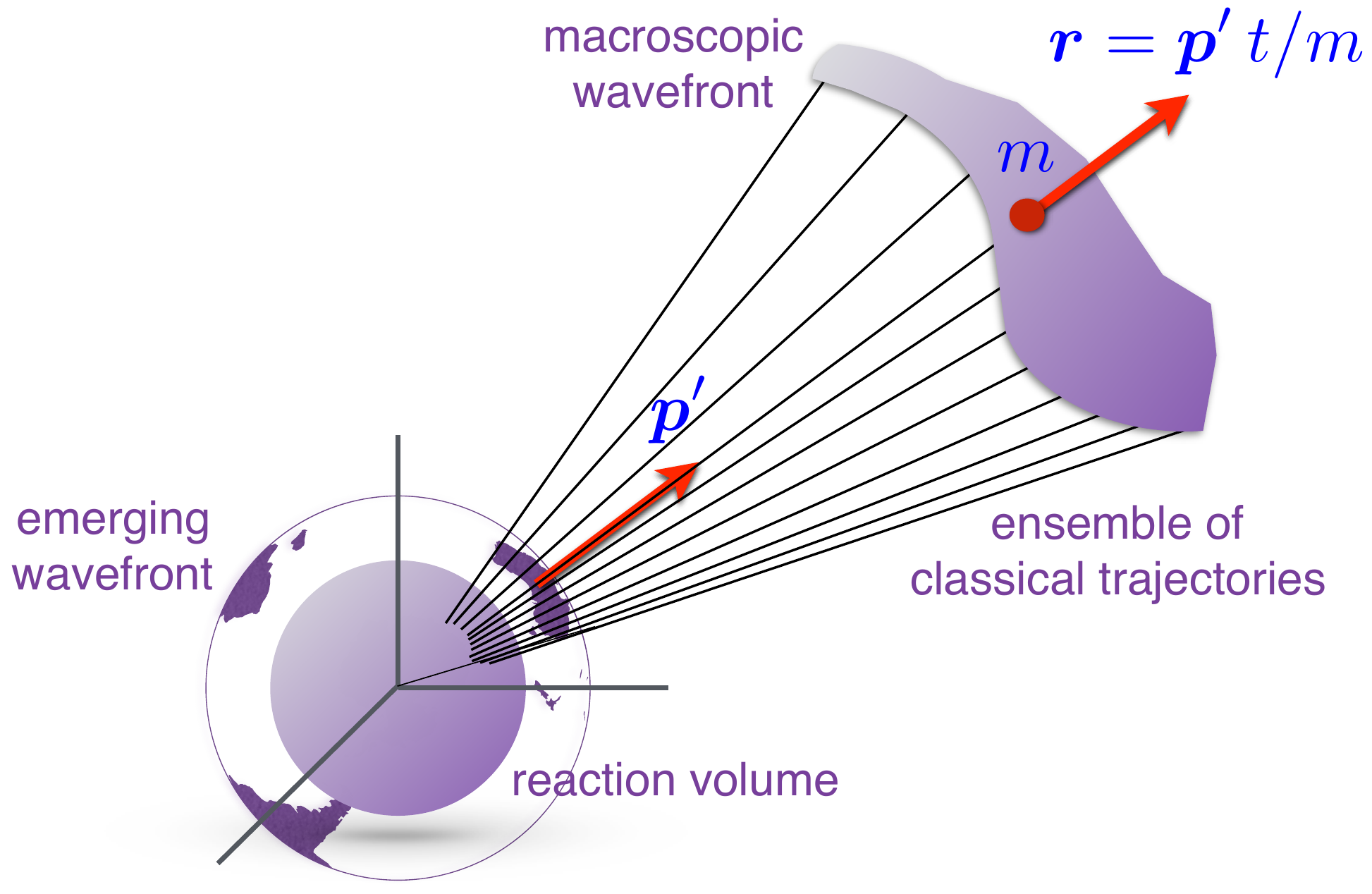}
\caption{\label{fig4} Illustration for free propagation of the spreading of the wave function with time along with the spreading of the underlying ensemble of classical trajectories. }  
\end{figure}

Here the spreading of the wave function with time, depicted in Fig.\ \ref{fig4}, is associated directly with the spread of the classical trajectories as the ensemble of particles moves out from the reaction volume according to $\vek r \approx (\vek p'/m)t$. 
The classical density factor $1/t^3$ has the interpretation that, as the swarm of trajectories moves out, the corresponding wave function
diminishes in amplitude to preserve normalisation.  

\subsection{Probabilities and particle-counting rates}
In an experiment where the microscopic interaction has a finite duration $\tau$, 
the arrival time $t$ of a detected scattered particle is the essential measured element. A detector at $\vek r$ will measure hits from repeated scattering events of scattered fragments along a particular classical trajectory at random times of arrival $t$. Thus one tracks and bins  the relative number of particles scattered into a solid angle $d\hat{\vek r} = d\Omega_{\vek r}$ incident on a detector area $dA_{\vek r} = r^2 d\Omega_{\vek r}$ at time $t$. The result corresponds to a probability
\begin{equation}
\label{dP}
dP = |\Psi(\vek r,t)|^2\,d \vek r
\end{equation}
integrated over the appropriate acceptance volume $d\vek r$ defined by the detector. Introducing the IT result \eref{locus2} one has immediately
\begin{equation}
\label{diffPp}
\frac{dP}{d\vek p'} \approx |\tilde\Psi(\vek p', \tau)|^2
\end{equation}
as the differential probability which can be connected to the relative number of particles exiting the reaction volume with momenta $\vek p'$ in the range $d\vek p'$ at $t=\tau$. The volume element $d\vek r$ defined by the detector defines also this volume element in momentum space according to the classical van Vleck trajectory connection \eref{VVdet}.

In the case of free motion of a particle of mass $m$, the IT \eref{kembleIT} along with \eref{freeVVdet} for $t \gg \tau$ gives
\begin{equation}
\label{phiITsq}
\frac{dP}{d\vek p'} \approx |\tilde \Psi(p' \hat{\vek r}, \tau)|^2 \approx (r/p')^3  |\Psi(\vek r, t \approx m r/p')|^2.
\end{equation}
The IT approximation becomes nearly exact for $\vek r$ and $t$ large enough and certainly for the macroscopic dimensions of a typical laboratory apparatus. 
Simulations of the IT connection \eref{phiITsq} are given in Sec.\ III of \rref{FeaginBriggs_ITIII} and in \rref{Macek}.

In this way, the spatial wave function at the detector is an image of the momentum wave function emanating from the collision.
(Of course, the images are extracted from the statistical ensemble of random arrival-time hits at the detector.) 
In time-dependent scattering theory, in particular necessary to describe short laser pulse interactions with matter, we show that this momentum wave function plays a fundamental role, corresponding to the role played by the scattering amplitude in time-independent scattering theory. The connection between the two is exposed in detail in section IV below.

\section{ Particle collisions and the initial channel in time-dependent theory}

Hitherto, attention has been confined to the outgoing channel and its description by a semiclassical wave function. This is appropriate for laser light interacting with an atomic or molecular target. However, 
generic collision experiments involve two sources of particles in the incident channel, either one beam impinging on a gaseous or solid target or two crossed particle beams.
 Hence, one should ask what is the role played by the IT in the incident channel, since here also particles traverse a macroscopic distance from the source to the collision region?
 
Many authors have expressed concern over the formulation of the initial conditions in quantum scattering theory. Particularly this 
concern is with the use of a plane wave $\psi_i(\vek r) \propto \exp{(i\vek p_i\cdot\vek r/\hbar)}$ as the incident wave function. For well-defined momentum such a wave occupies the whole of space which is plainly at odds with the tightly-collimated beam preparation necessary to achieve experimental accuracy. Also this  extended plane wave would interfere with the outgoing scattered wave. 

In more detailed expositions of scattering theory it is sought to remedy this deficiency. The standard approach (see for example \cite{GW,Newton,Joachain,Messiah}) is to represent a single incident particle by a narrow wavepacket. Then it is shown that under usual experimental conditions the spreading of the wavepacket is negligible and, using stationary-phase arguments, that the centroid of the wavepacket moves with a group velocity obeying classical mechanics. Similar non-spreading wavepackets are used to justify the form of the outgoing wave. 

Using such wavepackets it can then be shown, albeit in rather complicated fashion, that the same form of the differential cross section as given by an incident plane wave is obtained \cite{Taylor}.  As has been demonstrated, the IT obviates the need for wavepackets in the outgoing channel. The ensemble wave function with classical connection of position and momentum does not have to represent a single localised particle. In the IT the outgoing wave function spreads over all space and represents the {\emph{ensemble}} of  localised particle trajectories.

As we illustrate next for the incident beam, using the IT to derive a semiclassical wave function gives a  localised trajectory and a simpler explanation of why a plane-wave form for the incident wave is admissible. In the IT the wave function describes an ensemble of particles in directed classical trajectories emanating from a collimator.

\subsection{Time-dependent scattering theory}

In \eref{propfinal}, following the historical development of the IT, we considered only time propagation after the collision. To discuss the
complete scenario where particle detector {\emph{and source}} are at macroscopic distance from the collision region, we must describe the time development of the whole collision. Again, the time-dependent theory is appropriate to treat either time-dependent or time-independent potentials of interaction.
\begin{figure}[b]
\includegraphics[scale=.4]{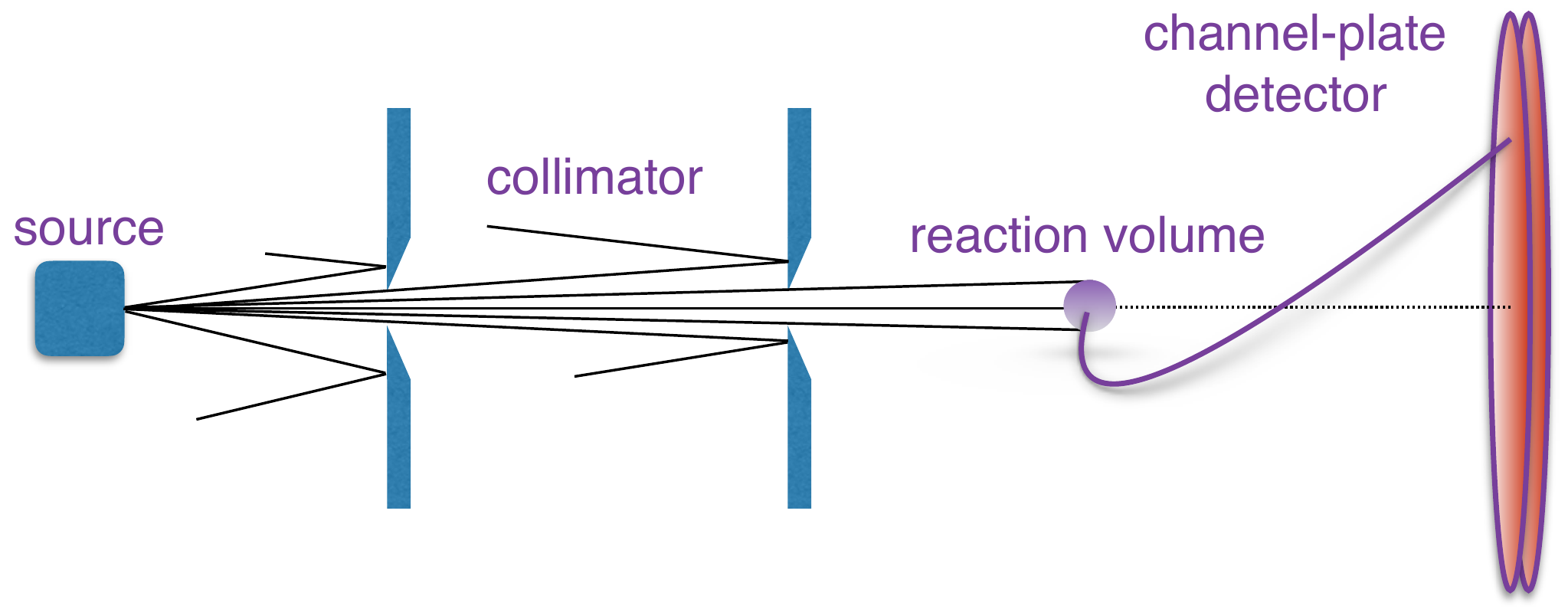}
\caption{\label{fig2} Illustration of a collision from source to detector of a particle described macroscopically by classical trajectories into and out of a microscopic quantum reaction zone. }  
\end{figure}

A collision of atomic particles can be viewed in three stages  as depicted in Fig.\ \ref{fig2}. First is the  preparation of a collimated beam and directing it
from a macroscopic distance onto a target area of microscopic dimensions. The particles emanating from the microscopic
collision volume move outwards over macroscopic distance before being registered by suitable detectors.

In the following, free motion of particles in stage one is assumed as this corresponds to the field-free motion from collimator to target. (That is, the incident beam either is so energetic or suitably shielded that the extraction fields can be ignored in stage one.)
By definition in stage two the potentials of interaction in the collision volume are of microscopic range and delineate the collision region. 

As before, we take $t=0$ as the time the collision begins.
In stage one the incident particles exit the collimator beam source at $t=-t_i$ and move freely for a time $t_i$ towards the collision region.  In stage two, the interaction potentials effectively switch on at $t=0$ and switch off at $t=\tau$, a short duration of atomic or molecular scale.
Of course, in principle this does not include the single most important interaction, the infinite range coulomb interaction of charged particles. For these interactions, the end of the collision region must be defined as the point at which the coulomb forces have diminished sufficiently that there is no further exchange of particles or excitations occurring. That is, the collision complex has ``frozen out" into its final channel wave function. Subsequent weak coulomb interaction between the outgoing particles is discussed in section VI.B and does not invalidate the IT.

 Stage three is when the particles move away from the collision region and are detected at a later time $t$. 
Since $t_i \gg \tau$ and $t \gg \tau$, the actions in both stages one and three are much greater than $\hbar$ and the motion can be approximated by the semiclassical time propagator,  i.e.\  one can use the IT result. Of course during the collision in stage two it is necessary to define the full quantum propagator. However, since the extraction fields in stage three are extremely weak compared to those operating during the collision, they can be ignored in stage two. 
  
The state at time $t$ is related to the initial state by the three-stage time propagator \cite{Greentime}, i.e.\ 
 \begin{equation}
 \begin{split}
 \label{3times}
 \ket{\Psi(t)} &= U(t,-t_i ) \ket{\psi_i} \\
&\approx U_F(t, \tau)\,U(\tau, 0)\,U_0(0, -t_i)\ket{\psi_i}, 
\end{split}
\end{equation}
where $\ket{\psi_i}$ is the time-independent  $t=-t_i$ initial state at the exit port of the collimator, $U_0$ is the propagator for free motion and $U_F$ that for motion in the extracting fields.
 
Then the free propagation from the collimator produces a state at time $t=0$ at the onset of the collision,
 \begin{equation}
 \label{inctimes}
\ket{\psi(0)} = U_0(0,-t_i)\ket{\psi_i} = \exp{\left[-\frac{i}{\hbar}H_0t_i\right]}\ket{\psi_i}.
\end{equation}

Accordingly, see \eref{Psiprop}, the  spatial wave function entering the collision region  is
   \begin{equation}
   \label{exact_inc}
 \psi( \vek R_i,0) = \int  \tilde K(\vek R_i, 0; \vek p, -t_i)  \, \tilde\psi_i(\vek p)\,d\vek p,
\end{equation}
where to facilitate introduction of the IT we have introduced $\vek R_i = \vek r_i' - \vek r_i$ to locate points $\vek r_i'$ in the reaction volume relative to the exit port of the collimator at fixed $\vek r_i$. (Recall the origin is at the centre of the reaction volume.)

Evaluated in SPA for $\vek r_i, t_i \rightarrow \infty$, this gives for the wave function incident on the target the IT result, see Eqs\ (\ref{IT_mom}) and  (\ref{freepropcl}),
   \begin{equation}
  \label{wfninc.}
 \psi(\vek R_i,0)  
  \approx   {\cal A}_i  \, \frac{e^{i \vek p_i\cdot \vek R_i/\hslash} }{(2\pi \hslash)^{3/2}} \, e^{-i p_i^2t_i/(2m\hslash)}
 \end{equation}
with semiclassical amplitude
   \begin{equation}
   \label{A_i}
  {\cal A}_i  
  \approx  (-i)^{3/2} (2\pi \hslash)^{3/2} \left(\frac{d\vek p_i}{d\vek r_i'}\right)^{1/2}\,  \tilde\psi_i(\vek p_i)  
 \end{equation}
along with the classical relation $\vek p_i =  m \vek R_i/t_i \approx -m \vek r_i/t_i$ from the SP condition.
This wave function decides the flux of particles incident and the final counting rate is proportional to this flux.

The incident trajectory bundle intersecting the microscopic reaction volume is characterized here by the single semiclassical trajectory along the collimator axis with momentum $\vek p_i$. 
The incident projectile plane wave described by $e^{i \vek p_i \cdot \vek R_i/\hslash}$ arises naturally near the reaction volume in \eref{wfninc.} as the semiclassical wavefront to this trajectory depicted in Fig.\ \ref{fig2}.
There is no need to introduce spatially-localised wavepackets constructed from an integral over momentum.

The collision takes place in the interval $0 < t < \tau$. Hence, from \eref{3times} and \eref{inctimes}, the state after the collision is
\begin{equation}
U(\tau, 0)\,U_0(0, -t_i)\ket{\psi_i} = U(\tau, 0)\,\ket{\psi(0)}.
\end{equation}
Projected into a mixed momentum-position space this gives an amplitude
\begin{equation}
\label{tzerowfn}
\begin{split}
\bra{\vek p}U(\tau, 0)&\ket{\psi(0)} \\
&= \int \bra{\vek p}\,U(\tau,0)\, \ket{\vek R_i}\,\psi(\vek R_i,0)\, d\vek r_i'. 
\end{split}
\end{equation}
Note the integral here is over the reaction volume coordinate $\vek r_i'$ as the position $\vek r_i$  of the collimator exit port is fixed.
If we recognise now that the amplitude ${\cal A}_i$ of incident trajectories in \eref{wfninc.}, for macroscopic $\vek r_i$, changes little over the variation of $\vek R_i$ (extent of $\vek r_i'$), then we can 
replace it by its value at $\vek r_i' = 0$. With this `plane-wave' approximation the integral in \eref{tzerowfn} can be performed to give
\begin{equation}
\label{pUppsi}
 \begin{split}
\bra{\vek p}U(\tau, 0)&\ket{\psi(0)} \approx \bra{\vek p}\,U(\tau, 0)\,\ket{\vek p_i} \,  {\cal A}_i  \, e^{-i p_i^2t_i/(2m\hslash)}  \\
	&= \tilde\Psi(\vek p, \tau) \, {\cal A}_i \, e^{-i p_i^2t_i/(2m\hslash)},
 \end{split}
\end{equation}
that is, the same momentum wave function as in \eref{Psiprop} but modified by the incident wave amplitude evaluated at $\vek r_i' = 0$.

The preparation of a collisional momentum wave function at the edge of the reaction volume is common to both time-dependent and constant interactions $V(t)$ and $V$, respectively.
This wave function propagates to asymptotic distances in the semiclassical approximation
which is our standard IT in the exit channel from section II.C. The state vector is
\begin{equation}
\ket{\Psi(t)} = U_F(t,\tau) U(\tau, 0)\,\ket{\psi(0)}
\end{equation}
with wave function, see \eref{Psiprop} and \eref{pUppsi},
\begin{equation}
\Psi(\vek r,t) = \int \tilde K_F(\vek r, t; \vek p, \tau)  \,  \tilde\Psi(\vek p, \tau)\ d\vek p \, {\cal A}_i \, e^{-i p_i^2t_i/(2m\hslash)}.
\end{equation}

The SPA approximation can be applied as before to give the result
   \begin{equation}
  \label{locus3}
|\Psi(\vek r,  t)|^2 =  \frac{d\vek p'}{d\vek r} \, |\tilde\Psi(\vek p',\tau)|^2 \,  \mathcal{P}_i,
\end{equation}
where we define the incident-wave probability density
\begin{equation}
\label{Pi}
\mathcal{P}_i \equiv |{\cal A}_i |^2 =  (2\pi \hslash)^{3}  \, \frac{d\vek p_i}{d\vek r_i'}  \, |\tilde\psi_i(\vek p_i)|^2
\end{equation}
evaluated at $\vek r_i' = 0$.

The only difference between \eref{locus3} and \eref{locus1}, applicable to strong laser excitation, is the additional factor $ \mathcal{P}_i$ from the incident particle beam. However, this term is simply proportional to the incident particle flux and so, of course, is the final particle counting rate.
 
The above time-dependent theory is applicable to any type of interaction with respect to time. The measurement is related to the scattered wave function in momentum space. However, for purely time-independent interactions the  time-independent theory is used usually. Then emphasis shifts to the definition of a scattering cross section in terms of a scattering amplitude, rather than a momentum wave function itself. The relation of our approach to time-independent theory, how a time-independent IT is formulated, and indeed how semiclassical quantum mechanics arises, is the subject of the next section.

\section{ The IT and time-independent scattering theory}

Time-independent scattering theory (TIST) is restricted to potentials of interaction which are themselves time-independent.  Since in most books on the subject, scattering theory  is presented in time-independent form, it is useful to connect the Kemble IT time-dependent limit to this standard formalism.

Therefore, in the next section of the paper, only time-independent interaction potentials are considered. These include both the potentials operating within the microscopic interaction volume and the much weaker extraction fields used in the asymptotic zone. As in section II, first the case of extraction fields is developed and then specialised to the simpler case of free motion.

\subsection{The time-independent IT limit}

Consider scattering into macroscopic and steady external fields, for example, an extraction electric field $\vek F$ in which the particle acquires a potential energy 
$V_F = - \vek F \cdot \vek r$. We emphasise that here the wave function is determined by  the external field $V_F$ of macroscopic extent beyond the short range of the \emph{microscopic} quantum scattering
potentials $V$. It is also important for the following to note that the extraction potentials are much smaller than the potentials operating during the scattering process, typically, $V_F \approx 10^{-8}\,V$. 
One can safely ignore $V_F$ inside the reaction volume.

Assuming a steady incident-beam current, the stationary-state $\ket{\Psi} $ describing the microscopic scattering and subsequent propagation to macroscopic distance in external fields is defined by the LS equation  \eref{LSGF} (ignoring the incoming wave)
 \begin{equation}
\ket{\Psi}  = G_F(E) V \, \ket{\Psi} =  \frac{1}{E - H_F + i \delta} \, V \, \ket{\Psi}.
\end{equation}
The Green operator$(E - H_F + i \delta)^{-1}$ defines the wave function in the final channel and so is defined by $V_F$ alone. 

The scattering wave function in coordinate space is defined accordingly  
 \begin{equation}
 \label{exactwfnc}
\Psi(\vek r,E)  = \int G_F(\vek r, \vek r'; E) \, V(\vek r') \, \Psi(\vek r',E) \,  d\vek r'.
\end{equation}

At this stage we replace $G_F(\vek r, \vek r'; E)$ formally with the semiclassical Green function, although for the cases of free motion or the uniform external extraction fields considered here, it turns out that the exact Green function is equal to its semiclassical approximation. 

 As Gutzwiller shows using the SPA \cite{Gutz}, the semiclassical Green function is of the form,
 \begin{equation}
 \label{Greenc}
G_F(\vek r, \vek r'; E)  \approx  \frac{1}{i \hslash} \frac{ \sqrt{D}}{2\pi i \hslash}  \,e^{i W(\vek r, \vek r'; E)/\hslash} 
\end{equation}
for an allowed classical trajectory connecting $\vek r'$ and $\vek r$ in the external fields. 
Now the phase function is precisely the time-independent classical action (Hamilton characteristic function) of \eref{TIcl}, defined as
 \begin{equation}
 \label{Greenc2}
 \begin{split}
W(\vek r, \vek r'; E) = \int_{\vek r'}^{\vek r} \vek p \cdot d\vek l,
\end{split}
\end{equation}
where $d\vek l$ is an element of path length along the trajectory.
The factor $D$ is given by
 \begin{equation}
 \label{D}
D  =  -\frac{\partial^2  W}{\partial E^2}  \frac{d \vek p'}{d \vek r}  
\end{equation}
and is the classical trajectory density but now along the fixed energy surface. 
This density can be expressed in a particularly useful form for characterising scattered trajectories. 
Berry and Mount (Eq.\  (7.42) p.\ 381) \cite{Berry} and Gutzwiller (Eq.\ (2.10) p.\ 25) \cite{Gutz} show that in the case of velocity-independent potentials 
 \begin{equation}
 \label{D_berry}
D  =  m^2 \frac{p'}{p}  \frac{d \Omega_{\vek p'}}{dA_{\vek r}},  
\end{equation}
where $d\Omega_{\vek p'}$ is the solid angle of a bundle of classical trajectories emerging in the direction of initial $\vek p'$  near $\vek r' \approx 0$ while $dA_{\vek r}$ is the cross section of the bundle when it reaches $\vek r$ with final momentum $\vek p$ due to the steering in the external fields.
That is, $d\vek p' = p'^2 dp' d\Omega_{\vek p'}$ while $d\vek r = dr \, dA_{\vek r}$ with $p' = \sqrt{2m[E - V_F(\vek r')]} \approx \sqrt{2m E}$ and $p = \sqrt{2m[E - V_F(\vek r)]}$.

For free-motion, straight-line trajectories emanating from the microscopic reaction volume, $dA_{\vek r} \approx r^2 d\Omega_{\vek p'}$ and $D \approx m^2/r^2$.
As shown below, this corresponds precisely to the $\sim \! 1/r^2$ fall off of the quantum probability density of the asymptotic scattered wave.

Invoking the scattering limit $r \gg r'$ and expanding the phase function around $\vek r' \approx 0$, one obtains
  \begin{equation}
 \label{W approx}
 \begin{split}
W(\vek r, \vek r'; E) &\approx 
W(\vek r, 0; E) + \frac{\partial  W}{\partial \vek r'} \Big|_{r'=0}\cdot  \vek r' \\
	&\equiv W(\vek r, 0; E) - \vek p' \cdot  \vek r',    
\end{split}
\end{equation}
where $\vek p'$ is the classical initial momentum just outside the reaction volume with $p' = \sqrt{2m[E - V_F(\vek r')]} \approx \sqrt{2m E}$.

Then, as a slowly-varying function of $\vek r' \approx 0$, the amplitude $D$ can be taken outside the integral
in \eref{exactwfnc} 
to relate the coordinate wave function to a scattering transition amplitude according to
 \begin{equation}
 \label{approxwfnc}
\Psi(\vek r,E)  \approx (2\pi \hslash)^{3/2} G_F(\vek r, 0; E) \,\bra{\vek p'}V \ket{\Psi},
\end{equation}
where now $\vek p'$ and $\vek r$ are connected by the classical trajectory in the external field. This equation is timeless and the classical mechanics also, corresponding to \eref{TIcl}.
The relation between $\vek p'$ and $\vek r$ is a phase-space connection 
 that distinguishes the various classical trajectories allowed for fixed $\vek r'$, $\vek r$, and $E$.
 Indeed \eref{approxwfnc} can be viewed as a time-independent form of the IT limit \cite{TI_Maslov}.

Although the quantum description embodied in \eref{approxwfnc} is time-independent, as usual for classical motion, it is convenient to introduce a time parameter. This corresponds to varying the energy $E$ and in particular a classical time of flight $T = t - \tau \equiv \partial W/\partial E$ along a trajectory from $\vek r'$ to $\vek r$ can be defined.
The factor $d\vek p'/d \vek r$ in \eref{D} is the same van Vleck factor defined in \eref{VVdet} but with the time replaced by $T = \partial W/\partial E$, so that $ \partial^2 W/\partial E^2 =  \partial T/\partial E$. 

Therefore one obtains a time-independent IT limit from \eref{IT_mom} or \eref{IT_coor} simply by dividing by the stationary-state time 
 factor $e^{-i Et/\hslash}$  and recognizing the Legendre transformation 
  $S(\vek r, t; \vek r', \tau) + E T  = W(\vek r, \vek r'; E) $ in the  $\vek r \gg \vek r'$   limit. 
That is,
\begin{equation}
\label{TIIT}
  \Psi(\vek r,E) \approx  i^{-3/2} \left(\frac{d \vek p'}{d \vek r} \right)^{1/2}  e^{i W(\vek r, 0; E)/\hslash} \,  \tilde\Psi(\vek p',E), 
\end{equation}
where now the van Vleck factor $d\vek p'/d \vek r$ is expressed as a time-independent phase-space relation.

With the introduction of the semiclassical Kemble IT limit, we have connected with the underlying scattered classical trajectory bundle. However, only relative time of flights are relevant, the instant of scattering or fragmentation is irrelevant and in fact unknown. As is shown below, the familiar spreading of the classical trajectory bundle with time, for example $\vek r \sim (\vek p'/m) t$ for free motion, corresponds precisely to the $\sim \! 1/r^2$ fall off of the quantum probability density.

\subsection{The relation between scattering amplitude and momentum wave function}

In time-independent scattering theory it is customary to define the scattering amplitude as
\begin{equation}
\label{fp}
f(\vek p') = -\sqrt{\frac{2\pi}{\hslash}} m \,\bra{\vek p'}V \ket{\Psi}. 
\end{equation}

Then from \eref{approxwfnc}, we obtain
\begin{equation}
\label{ITfsq}
| \Psi(\vek r,E)|^2  \approx m^{-2} D \, |f(\vek p')|^2.  
\end{equation}
For the special case of free-motion outside the reaction volume, the factor $m^{-2} D$ is given approximately by $r^{-2}$, and we recover the familiar relation of the asymptotic coordinate wave function with the scattering amplitude (see \eref{asymp} below).

From (\ref{TIIT}) along with the density from \eref{D}, one also has a connection between the  scattering amplitude and the momentum scattering wave function in the form
\begin{equation}
\label{fsq}
|f(\vek p')|^2 \approx m^2 D^{-1} \, \frac{d \vek p'}{d \vek r}  |\tilde\Psi(\vek p')|^2 \approx m^2\,  \left| \frac{\partial T}{\partial E} \right|^{-1}\, |  \tilde\Psi(\vek p',E)|^2.
\end{equation}

Generally, however, in the presence of extraction fields, we obtain with \eref{D_berry} the detection probability in a short interval $dt$ (relevant to steady incident beams) as
\begin{equation}
\label{dP}
 | \Psi(\vek r,E)|^2 \, d\vek r \approx | \tilde \Psi(\vek p',E)|^2 \, d\vek p' \approx  \frac{p'}{m} dt  \, |f(\vek p')|^2 \, d\Omega_{\vek p'},
\end{equation}
where we have inserted $dr = (p/m) dt$ so that $(p'/p) dr = (p'/m) dt = dr'$.

In the Gutzwiller notation for the densities in time $C(t) \equiv d \vek p'/d \vek r $ and in energy $D(E)$ 
from Eqs.\ (\ref{Krr'}) and (\ref{Greenc}), we have from \eref{ITfsq} the equality
\begin{equation}
\label{CD}
| \Psi(\vek r,E)|^2 \approx C(t) | \tilde\Psi(\vek p',E)  |^2  \approx m^{-2} D(E) \, |f(\vek p')|^2.  
\end{equation}

If instead we introduce the closely related $\cal T$-matrix element as
\begin{equation}
{\cal T}(\vek p') = \sqrt{\frac{2\pi}{\hslash}} \,\bra{\vek p'}V \ket{\Psi}, 
\end{equation}
then we can write the asymptotic relation \eref{CD} in the form
\begin{equation}
 C(t) | \tilde\Psi(\vek p',E)  |^2  \approx  D(E) \, |{\cal T}(\vek p')|^2.  
\end{equation}

This general result for the asymptotic wave function is new and provides a connection between the time and energy representations of scattering theory for time-independent interactions. It also underscores Gutzwiller's emphasis on the relation between the time and energy classical density factors 

Relations between the scattering amplitude and its semiclassical equivalent have been given before \cite{Pech1,RostHeller} but where the {\emph{complete}} scattering event is treated semiclassically. We emphasise here that in \eref{fsq} the amplitude and momentum wave function are those arising from a full quantum mechanical treatment within the reaction volume and the relations Eqs.\ (\ref{ITfsq}) and (\ref{dP}) emphasise that only outside the reaction volume is the motion semiclassical.

 \subsection{The time-independent IT limit for free asymptotic motion}

 As before, it is illustrative to work through the case of free motion in detail. The key element of time-independent IT limit is the time-independent classical action.
 From \eref{Greenc2} with $V_F = 0$ one has
\begin{equation}
 \label{W0}
W(\vek r, \vek r'; E) = \sqrt{2m E} \, R = p' R,  
\end{equation}
where $\vek R = \vek r - \vek r'$ and $p' = \sqrt{2m E}$. 

As required, the magnitude of the final momentum is equal to that of the initial momentum,
\begin{equation}
 \label{pp'0}
\begin{split}
\vek p &= \partial W/\partial \vek r = p' \hat{ \vek R}, \\
\vek p' &= -\partial W/\partial \vek r' = p' \hat{ \vek R}, 
\end{split}
\end{equation}
 along the straight-line free trajectory $\vek R$.

The classical time of flight from $\vek r'$ to $\vek r$ for fixed energy is given by
 \begin{equation}
 \label{t0}
\begin{split}
T = \frac{\partial W}{\partial E} &= \sqrt{\frac{m R^2}{2E}} = \frac{m R}{p'}, \\
\frac{\partial T}{\partial E} &= -\frac{m^2 R}{p'^3} = -\frac{T}{2E}.  
\end{split}
\end{equation}
Then one obtains for the van Vleck factor \eref{freeVVdet} the further relation, 
\begin{equation}
 \label{VVdet0}
\frac{d\vek p'}{d \vek r} = \left(\frac{m}{T}\right)^3 =  \left ( \frac{p'}{R} \right)^3,
\end{equation}
which is the time-independent form to be used in the IT limit of \eref{TIIT}.

The Green function density factor \eref{D} is then simply $D = m^2/R^2$,
so that substitution of this result along with \eref{W0} in \eref{Greenc} gives
 \begin{equation}
\label{G0}
G_0(\vek r, \vek r'; E) = -\frac{m}{2\pi\hbar^2}\frac{e^{i p' |\vek r - \vek r'|/\hslash}}{|\vek r - \vek r'|},
\end{equation}
which is the exact free Green function and merely demonstrates that the semiclassical approximation for free motion is exact.

The above relations are of course valid also in the asymptotic limit $r \gg r'$. 
 In this limit, $\vek p' \approx p' \hat{ \vek r}$ and
one derives from Eqs.\ (\ref{approxwfnc}) and (\ref{fp}) the usual asymptotic form of the scattering wave function in TI theory,
\begin{equation}
\label{asymp}
 \Psi(\vek r, E) \approx \frac{e^{i p' r/\hslash}}{r} f(\vek p').
\end{equation}
Also, directly from \eref{ITfsq} with $m^{-2} D \approx r^{-2}$, one obtains
\begin{equation}
\label{newresult}
 |\Psi(\vek r,E)|^2 \approx \frac{1}{r^2} |f(\vek p')|^2 \approx  \frac{d\vek p'}{d\vek r} \, |\tilde\Psi(\vek p', E)|^2.
\end{equation}

These relations all illustrate the fact that even in time-independent scattering the asymptotic motion is semiclassical, the wave function coordinates follow classical trajectories, and the scattering amplitude is related to the momentum-space wave function resulting from the collision.

  \subsection{Scattering cross section}

 For time-independent interactions during the collision, it is usual and meaningful to define a cross section. Again we consider  semiclassical propagation starting from the exit port of a distant collimator into the reaction volume, where the microscopic quantum scattering takes place, followed by semiclassical propagation out to a distant detector as depicted in Fig.\ \ref{fig2}. 
 Although we confine the derivation here to elastic scattering, the results readily generalise to fragmentation and rearrangement reactions.

We begin with the \eref{LSG} but again for just the scattered wave 
\begin{equation}
 \label{LippSchw2}
\ket{\Psi}  = G \, V \, \ket{\psi},
\end{equation}
where we assume as in section III.A the incident beam described by $\ket{\psi}$ is suitably shielded from the external extracting fields.  

Using the identity
\begin{equation}
 \label{G}
G = G_F + G_F V G
\end{equation}
in \eref{LippSchw2}
gives
\begin{equation}
 \label{T'}
\begin{split}
\ket{\Psi}  &= G_F V(1 + G V)\ket{\psi} \\  &  \equiv G_F {\cal T}  \ket{\psi},
\end{split}
\end{equation}
where $ {\cal T} = V + VGV$ is the transition operator, which defines the scattering amplitude.
Expressing this result in the coordinate representation using $\vek R_i = \vek r_i' - \vek r_i$ to facilitate introducing the semiclassical approximation, see \eref{exact_inc}, gives
\begin{equation}
 \label{Psi_c}
\begin{split}
\Psi(\vek r, E) \approx  \iint d\vek r'  \,  d\vek r_i' \, G_F(\vek r, \vek r'; E) \bra{\vek r'} {\cal T} \ket{\vek R_i}  \psi(\vek R_i, E)
\end{split}
 \end{equation}
describing the collision in the three stages stages of particle propagation, from a distant collimator port at fixed $\vek r_i$ into and across the reaction volume defined by  $\vek r_i', \vek r'$ followed by propagation to a distant detector port at $\vek r$. 
With the relative separation of the two colliding particles as coordinate, the origin is at the centre of the reaction volume. Then the range of $V$ and therefore of ${\cal T}$ limits $\vek r_i'$ and $\vek r'$ to microscopic dimensions compared to $\vek r_i$ and $\vek r$.

 Now following the procedure we used in section III.A to obtain \eref{TIIT},  the IT approximation from \eref{wfninc.}  gives for the \emph{time-independent} incident wave 
   \begin{equation}
 \psi(\vek R_i,E)  
  \approx   {\cal A}_i  \, \frac{e^{i \vek p_i\cdot \vek R_i/\hslash} }{(2\pi \hslash)^{3/2}} 
 \end{equation}
with semiclassical amplitude $ {\cal A}_i$ given in \eref{A_i} describing the IT density of the incident trajectory bundle exiting the collimator at $\vek r_i$. 

With the incident beam suitably shielded from external fields, we have that $\vek p_i \approx -p_i \hat{\vek r}_i$ with $p_i = \sqrt{2m E}$ and $E$ the energy of the incident beam.

Also we introduce the semiclassical approximation \eref{Greenc} for the outgoing Green function $G_F$.
Then, expanding the slowly-varying functions  $D$ and $W$ in terms of small $\vek r'$, one obtains
\begin{equation}
 \label{GF_IT}
G_F(\vek r, \vek r'; E) \approx (2\pi \hslash)^{3/2} G_F(\vek r, 0; E) \, \frac{e^{-i \vek p' \cdot \vek r'/\hslash} }{(2\pi \hslash)^{3/2}}, 
\end{equation}
where (cf.\ \eref{W approx})
\begin{equation}
\vek p' = -\frac{\partial  W(\vek r, \vek r'; E) }{\partial \vek r'} \Big|_{r'=0}
\end{equation}
is the classical momentum at the outer limit of the reaction volume.
The $\vek r'$ and $\vek r_i'$ integrals in \eref{Psi_c} can now be performed and we obtain 
\begin{equation}
 \label{Psi_IT}
\Psi(\vek r, E) \approx   
	(2\pi \hslash)^3 \, G_F(\vek r, 0; E)  \,  \bra{\vek p'} {\cal T} \ket{\vek p_i} \, {\cal A}_i.
\end{equation}

One can define a state $\Psi_i$ within the microscopic scattering volume developing from the momentum eigenstate $ \ket{\vek p_i}$ as $\ket{\Psi_i}  \equiv (1 + G V)\ket{\vek p_i}$. 
Thereby one obtains the transition matrix element describing the microscopic scattering in the standard form, 
\begin{equation}
 \label{Trep}
\bra{\vek p' } {\cal T} \ket{\vek p_i} = \bra{\vek p'} V(1 + GV) \ket{\vek p_i} \equiv  \bra{\vek p' } V \ket{\Psi_i},
\end{equation}
so that
\begin{equation}
 \label{Psi_IT_fscatter}
\Psi(\vek r, E) \approx   
	(2\pi \hslash)^3 \, G_F(\vek r, 0; E)  \,  \bra{\vek p' } V \ket{\Psi_i}  \,  {\cal A}_i,
\end{equation}
which is the analog of \eref{approxwfnc} but now including the incident wave amplitude $ {\cal A}_i$.

Assuming elastic scattering one has  $ p' = p_i$ 
so that  the density of the scattered trajectory bundle from \eref{D_berry} is given by
\begin{equation}
 \label{Df}
D \approx m^2 \frac{p_i}{p} \frac{d\Omega_{\vek p'}}{dA_{\vek r}},
\end{equation}
where $d\Omega_{\vek p'}$ is the solid angle of asymptotic trajectories just emerging from the reaction volume along $\vek p'$ near $\vek r' \approx 0$ while $dA_{\vek r}$ is the cross-sectional area of the `externally steered' bundle arriving at a distant detector port at $\vek r$. For free macroscopic propagation without extraction fields, $p = p_i$ and $d\Omega_{\vek p'}/dA_{\vek r} \approx 1/r^2$.

Collecting results, one generalizes the scattered spherical wave \eref{asymp} to include external field extraction semiclassically with
\begin{equation}
 \label{Psi_IT2}
\Psi(\vek r, E) \approx   
	\sqrt{ \frac{p_i}{p} \frac{d\Omega_{\vek p'}}{dA_{\vek r}} } \; e^{i  W(\vek r,0; E) /\hslash}  f(\vek p', \vek p_i) {\cal A}_i, 
\end{equation}
where the scattering amplitude is defined by
 \begin{equation}
 \label{fscatt}
f(\vek p', \vek p_i) \equiv -\sqrt{\frac{2\pi}{\hslash}} m \,   \bra{\vek p' } V \ket{\Psi_i} (2\pi \hslash)^{3/2}.
\end{equation}

The macroscopic scattered wave \eref{Psi_IT2} defines a differential scattering probability into a volume element $d\vek r = dr \, dA_{\vek r}$ in a short time interval $dt$ according to (cf.\ \eref{dP})
 \begin{equation}
 \label{detectionP}
dP  \approx  |\Psi(\vek r, E)|^2 \, dr \, dA_{\vek r} \approx  {\cal P}_i   \, \frac{p_i}{m} dt   \, |  f(\vek p', \vek p_i') |^2 d\Omega_{\vek p'},
\end{equation}
where  ${\cal P}_i = |{\cal A}_i|^2$ from  \eref{Pi}.
We then define a cross section for the microscopic scattering reaction as the effective area $d\sigma$ the exit channel defined by $d\Omega_{\vek p'}$ presents in the time $dt$ to a steady incident beam with speed $p_i/m$ and density $\mathcal{P}_i$. That is, $ \mathcal{P}_i \, (p_i/m) dt \, d\sigma \equiv dP$. Thus one obtains the differential scattering cross section as
 \begin{equation}
 \label{xsec2}
\frac{d\sigma}{d\Omega_{\vek p'}} = |  f(\vek p', \vek p_i') |^2,
\end{equation}
independently of the external fields.

Measurement of the cross section requires the detection counting rate at $\vek r$ relative to the incident beam current along with a determination of the classical ballistic trajectory defined by $\vek r$ and the \emph{direction} of the scattered momentum $\vek p'$ exiting the reaction volume near $\vek r' \approx 0$. For elastic scattering, the magnitude of $\vek p'$ is already fixed by the incident energy, $p' = p_i = \sqrt{2m E}$.

\subsection{Measurement of time delays}

Generalized to $N$ scattered particles and a set of $N$ fixed detector positions $\vek r$, the initial momentum distribution $ |\tilde\Psi(\vek  p',E)|^2$ can be inferred from a detector hit distribution $|\Psi(\vek r,E)|^2$ (cf.\ \eref{phiITsq}). One only requires the classical trajectory density generalized to $N$ particles as in \eref{VVnFree} for free propagation but expressed in time-independent form using \eref{VVdet0} as 
\begin{equation}
 \label{VVnFreeTI}
 \frac{d \vek p'}{d\vek r}  =   \prod_n  \left ( \frac{m_n}{T_n} \right)^3 \approx  \prod_n   \left ( \frac{p_n'}{r_n} \right)^3 
\end{equation}
for $r_n \gg r_n'$. 

There is a caveat, however. When the total energy is fixed as in the case of a steady incident beam, the instant of fragmentation is irrelevant (and in fact unknown) and individual particle time of flights $T_n \approx m_n r_n/p_n'$ cannot be measured.
Instead, the initial momenta $p_n'$ can be determined in terms of $N-1$ measured \emph{time delays} $\Delta T_n = T_n - T_1$ between detection at say detector $n=1$ and the remaining detections. In particular, one can solve the $N$ equations
\begin{equation}
 \label{delays}
\Delta T_n \equiv  \frac{m_n r_n}{p_n'} -  \frac{m_1 r_1}{p_1'}, \hspace{.1in} E = \sum_n \frac{p_n'^2}{2m_n},
\end{equation}
for the $p_n'$ in terms of the delays $\Delta T_n$, the detector positions $r_n$, and $E$.

Although in general the solution must be obtained numerically, for two-particle detection  
as in the low-energy $(e, 2e)$ experiment of Cvejanovic and Read \cite{Read}, an analytical solution can be given.

Assume for simplicity identical particles $m_1 = m_2 = m$ and detectors equally spaced from the origin $r_{1} = r_{2} = r$ and introduce a dimensionless delay time $ \tau \equiv \Delta T/t$ with $t \equiv \sqrt{m r^2/E}$ from \eref{t_W}. Then one finds that 
\begin{equation}
 \label{p1p2}
p_n'=  \sqrt{m E} \, \frac{\mp 1 \pm \sqrt{1+2\tau^2} + \sqrt{2} \sqrt{\tau^2 + \sqrt{1+2\tau^2} -1}}{2\tau}
\end{equation}
with the upper $\pm$ signs for $p_1'$ and the lower for $p_2'$. 

In the limit $\Delta T \rightarrow 0$, one sees that $p_1' \rightarrow p_2'$ with $E_{n=1,2}' \rightarrow \small{\frac{1}{2}} E$,  as desired. Moreover, as $\Delta T  \rightarrow +\infty$, one verifies that $p_1' \rightarrow \sqrt{2m E}$ and $p_2' \rightarrow 0$,  so that $E_1' \rightarrow E$ and $E_2' \rightarrow 0$, and vice versa as $\Delta T  \rightarrow -\infty$.
\begin{figure}[b]
\includegraphics[scale=.22]{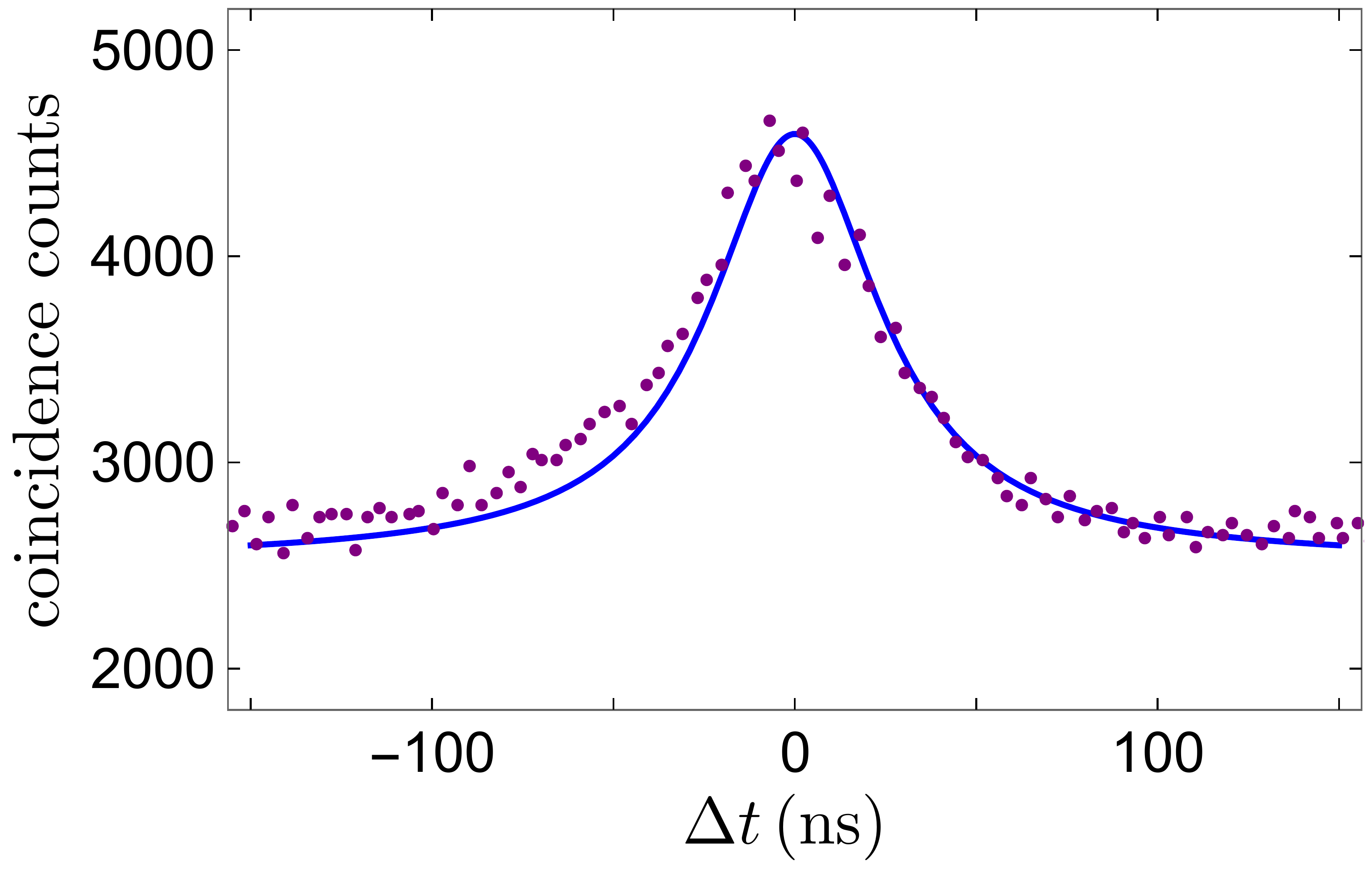}
\caption{\label{fig3}  Electron-pair coincidence data points (purple dots) from \cite{Read} for back-to-back ejection $\vek r_1 = -\vek r_2$ as a function of the time delay $\Delta T$ between detections for each pair ejected. The solid (blue) curve is a fit based on the macroscopic IT wave function \eref{PsisqLab_N_IT} using Eqs.\  (\ref{p1p2}) and (\ref{Psip1p2}) with $\sigma = 1 \, \mbox{a.u.}$  and $\Sigma = 10 \, \mbox{a.u.}$ arbitrarily normalized. }  
\end{figure}

To model the initial momentum wave function of the ionized electron pair, consider a simple product of gaussians
\begin{equation}
 \label{Psip1p2}
 \tilde\Psi(\vek  p_1', \vek  p_2', E) \equiv \tilde\psi_{\sigma}(\vek  p') \, \tilde\psi_{\Sigma}(\vek  P')  
\end{equation}
describing the relative $\vek p'$ and center of mass $\vek P'$ momentum distributions of the pair with widths $\sigma$ and $\Sigma$, respectively. 
Here $\vek p' = \frac{1}{2} (\vek  p_1' - \vek  p_2')$ and $ \vek P' = \vek  p_1' + \vek  p_2'$, so 
$\tilde\Psi(\vek  p_1', \vek  p_2', E) $ is not a simple product of $\vek  p_1'$ and $\vek  p_2'$ functions  
and $\Psi(\vek r_1, \vek r_2,E)$ from \eref{PsisqLab_N_IT} is likewise entangled as a function of $\vek r_1$ and $\vek r_2$. 

Fig.\ (\ref{fig3}) shows the $E = 0.37 \, \mbox{eV}$  coincidence data from \cite{Read} as a function of $\Delta T$ from a pair of detectors arranged for back-to-back ejection $\hat{\vek r}_1 = -\hat{\vek r}_2$ at equal distances $r_1 = r_2 = 2 \, \mbox{cm}$ from the reaction volume. The fit is the \emph{macroscopic} coordinate distribution from \eref{PsisqLab_N_IT} calculated using Eqs.\  (\ref{p1p2}) and (\ref{Psip1p2}) with $\sigma = 1 \, \mbox{a.u.}$  and $\Sigma = 10 \, \mbox{a.u.}$
Although normalized arbitrarily, the simple model of electron-pair correlations embodied in \eref{Psip1p2} describes the data well.

\section {Commentary on the IT}

In this section we comment on the application of the IT to situations where there may be interactions in the asymptotic zone, not only due to charged particles moving in
applied external fields but due to interactions between the particles themselves. In addition we point out that typical quantum effects such as interference patterns and state entanglement are preserved by the IT. There are many studies of classical and semiclassical approximations for the description of continuum particle motion following fragmentation by laser or particle beams. Essentially they are all related to the Kemble IT limit studied in this paper.

\subsection{Attosecond physics and classical motion in the continuum}

We have considered the application of the IT to both time-dependent and time-independent collision theory. Although Kemble's derivation dates back to 1937, the IT has been applied relatively recently to time-dependent interactions during ion-atom collisions \cite{Macek}. More modern emphasis is placed on ionisation in short intense laser pulses, which clearly require a time-dependent approach. Indeed, one aim of such experiments is often to track the motion of ionised electrons on the attosecond time scale. Although not mentioned, the IT plainly is implicit in such considerations. A prominent example is the theory of re-scattering of continuum electrons from the oscillating classical electric field of the laser. This results in electrons returning to the parent positive ion and re-combining leading to the generation of high harmonics of the field \cite{Lew}. This is described fully quantum mechanically but also with classical electron motion \cite{Kul}. However, quite generally, such continuum motion of electrons arising from field ionisation is described routinely semiclassically or wholly classically, even for slow electrons \cite{Rost2}. The IT is ultimately the justification for this.

These experiments usually concern one electron in the continuum but the IT applies also to several particles in the final channel, as we discuss next.

\subsection{ Particle interactions in the continuum}

Where composite particles are emitted from the reaction volume, their internal state continues to require a quantum description.  The same applies to intrinsic particle spin of course. The IT has been derived above for the continuum motion of particles possibly extracted by external fields, which automatically implies electrically-charged particles. 
 If more than one charged particle is emitted from a single collision event, in principle they may interact. However, the particles are usually so well separated in phase space (position and/or momentum) that the Coulomb interaction is negligible after the particles have separated to macroscopic distances. In any case, if not negligible, the interaction in Coulomb fields is easily treated in classical mechanics.
 
In extreme cases the correlation due to long-range Coulomb interaction may persist out to the detector. This can occur when the relative momentum of the two particles is small.
 A famous example of this is the phenomenon of ``Coulomb capture to the continuum" \cite{Macek2}. Here a positively-charged ion or a positron ``captures" an initially-bound electron
and the two free particles move to the detector, essentially in a low relative-momentum continuuum state. Usually the pair, although slightly unbound, are considered as a composite particle and their relative motion treated appropriately by quantum mechanics. 

The same phenomenon can occur when two fast electrons move out of the collision volume with low relative momentum and there is a strong repulsion between them \cite{Preben}.
Also, the so-called ``post-collision interaction" occurs between two  electrons emitted at different times from a collision complex. 

An even more famous example of the Coulomb correlation persisting beyond the confines of the strong-interaction collision volume is the emission of two electrons just above the double-ionisation threshold, the ``Wannier" phenomenon. Here two electrons move in their mutual field and that of the residual positive ion \cite{Peterkop, Feagin, Rost3}. Interestingly, Wannier's original analysis assumed classical motion at all distances and this use has been justified \cite{Rost3}.  

One also sees that, even in the extreme case of a very slowly-moving light particle, let us say that of a milli-eV electron, the conditions for validity of the IT, i.e.\ accumulated action many times greater than $\hslash$, are readily satisfied.  A 1 meV electron has energy of approximately $10^{-5}$ in atomic units (a.u.) and therefore after a time of  the order of $ 10^6 \, \mbox{a.u.}$, that is after $10^{-11}  \, \mbox{sec}$, the action has reached some $10 \, \hslash$. This corresponds to a distance travelled of  roughly $ 3000 \, \mbox{a.u.}$  from the collision volume, which is still microscopic compared to typical detector distances of centimetres from the collision volume.

Thus, in all practical cases charged particles offer no essential difficulty compared to the motion of neutral particles. 

\subsection{Quantum interference and entanglement}

Considering first a single charged particle, the IT predicts a contribution to the ensemble wave function from each well-defined classical trajectory in the external field. 
Each trajectory and contribution to the asymptotic wave function is unique, so that if a measurement specifies the full vector coordinates there is no interference of contributions. This corresponds to a ``which way" measurement. However, often the determination is imprecise. For example a spherical analyser would detect only scalar distance from the reaction volume. Then interference patterns are created by ensembles of identical particles. 

It may happen that the geometry of the extracting fields supports more than one trajectory reaching the same detection point $\vek r$ but with differing initial momentum directions $\vek p'$ exiting the reaction volume resulting in an interference of the semiclassical wavefronts along the different trajectories.  A classic example is the \emph{photodetachment microscope} experiment of Blondel and coworkers \cite{Blondel} involving two intersecting trajectories with different times of flight (TOF) and for fixed energy. As explained theoretically in \rref{Manfred} interference is observed for photo-electrons extracted by a constant electric field, where electrons moving directly to the detector interfere with electrons ejected in the opposite direction and then turned around by the field. 
There are more recent realisations  of this phenomenon such as \emph{velocity map imaging} (VMI)  \cite{BordasRobicheaux}. 
Another example is the atom interferometer \cite{BriggsFeaginNJP} where intersecting and interfering trajectories  have the same TOF.
Note, in these cases, a measurement of final momentum \emph{direction} $\vek p$ would distinguish the different trajectories and eliminate the interference.

The interpretation of quantum interference patterns arising from identical particles has been analysed in great detail in the semiclassical trajectory approximation, both using time-independent and time-dependent descriptions of the transition from reaction volume to detector. Details are to be found particularly in the papers of Kleber and co-workers \cite{Manfred}.
 
That the entanglement of many-particle wave functions, encoded in the momentum wave function at the edge of the reaction volume, is preserved and imaged out to macroscopic
distances and times is clear from the  many-particle IT of \eref{AgainIT}. Coincident detection of particles from the same microscopic scattering will register all features arising from entanglement.

The occurrence of  nodes, either in single-particle or in many-particle wave functions, also presents no difficulty in the IT. The nodes correspond simply to the absence of a particular
trajectory in the classical ensemble represented by $d\vek p'/d\vek r$, which is traced to the vanishing of 
$\tilde\Psi(\vek p', \tau)$  for that value of $\vek p'$.

\section{Conclusions}

We have presented a re-formulation of both time-dependent and time-independent scattering theory based upon a semiclassical representation of the asymptotic quantum wave function in both the
incident and final channels. This semiclassical approximation, valid in the large time, or large distance limit generalises the Kemble imaging theorem limit, applied hitherto to free asymptotic motion.

The semiclassical limit justifies the standard use by experimenters of classical mechanics to describe the field extraction of scattered fragments but also, since the quantum wave function remains intact out to macroscopic detector position, explains how typical quantum effects such as interference and entanglement are preserved.

In the incident channel the IT limit, which gives a classical relation between position and momentum quantum coordinates, leads to the emergence of a directed plane wave form with fixed momentum
and position coordinate limited by proportionality to this fixed value. Previously, conceptual difficulties with the use of a completely de-localised plane wave have led to somewhat complicated scattering treatments involving narrow wavepackets in the incident channel. The directed nature of the IT result, in which the plane wave arises naturally from an expansion of the classical action, render such treatments unnecessary.

 Traditionally the IT has been derived for a time-dependent asymptotic final-channel wave function and applied to interpret the results of time-dependent theory, particularly from laser and heavy-ion interactions with atoms \cite{Macek,Groz}. Here we have shown that time-independent scattering theory can be described asymptotically also by a semiclassical wave function. This leads to a time-independent version of the IT. In particular we have been able to connect the scattering wave function in momentum space, the main collision property arising from the IT, with the usual definition of a scattering amplitude or momentum-space T-matrix element, the main collision property of standard scattering theory. Interestingly, the connection between these two quantum elements is manifest in terms of two {\emph{classical}} trajectory-density factors defined by Gutzwiller \cite{Gutz}.

  The time-independent version of our approach illuminates the text-book derivation of scattering theory in that it exposes the semiclassical mechanics implicit in the asymptotic procedure of that theory. In addition, we have shown that  the semiclassical treatment leads to the usual formula for a cross section, irrespective of whether extraction fields are used or not. Furthermore, this has allowed us to present the generalised IT applicable to arbitrary many-particle collisions.

\appendix

\section{Equations of motion}

\subsection{ Time and energy in mechanics }

The time-independent form of classical mechanics leads to the concept of structures in position-momentum $(\vek r, \vek p)$ time-independent phase space. The potentials of interaction are considered time independent and usually a function of $\vek r$ alone. It is little used for calculation since time-parametrised versions such as Newton's equations are more practical. Nevertheless phase space structures are vital to the understanding of features of regular and irregular parts of phase space, as in chaotic motion for example.

In fixed-energy representation, the total energy  $E$ is conserved and the basic equation, $H = E$, where $H$ is the total Hamiltonian expresses this. One method to solve the dynamics is to write the fundamental equation as the Hamilton-Jacobi differential equation in terms of the classical action function $W$ (Hamilton's characteristic function), i.e.
\begin{equation}
H(\vek r,\vek p) - E = 0,
\end{equation}
or
\begin{equation}
\label{TIcl}
H(\vek r, \nabla W) - E =0.
\end{equation}

For interactions not explicitly dependent on time, the time version introduces parametrisation with an external time to give $\vek r(t), \vek p(t)$ and a new action, Hamilton's principal function 
$ S = W - Et$, such that the Hamilton-Jacobi differential equation becomes 
\begin{equation}
\label{TDcl}
 \left(H(\vek r, \nabla) + \frac{\partial }{\partial t}\right)S(t) = 0.
\end{equation}
For explicitly time-dependent interactions only this latter equation can be used and the action function $S$ must be changed appropriately.

The quantum versions of the dynamic differential equations are rather similar. In the time-independent case one has the TISE, the analogue of \eref{TIcl}
\begin{equation}
[H(\vek r, -i\hbar\nabla) - E]\psi =0,
\end{equation}
where $\psi$ is the wave function. The time-dependent version is of course the TDSE, the analogue of \eref{TDcl}
\begin{equation}
\left(H(\vek r, -i\hbar\nabla) -i\hbar \frac{\partial }{\partial t}\right)\psi(t) = 0.
\end{equation}
Again, for time-dependent interaction potentials only this form can be used. However, it is to be noted that in the quantum equations, time enters the interactions explicitly only when a classical approximation of (quantised) external fields is used. Then the hamiltonian $H$ becomes parametrically dependent upon time. A prime example is the use of a classical solution of Maxwell's equations to describe a strong laser field, rather than quantising the electromagnetic field.

The connection of quantum and classical descriptions through the semiclassical wave function is sketched as follows. Quite generally one can write the  wave function as 
 \begin{equation}
\psi(\vek r,t) = e^{i \Phi(\vek r,t)/\hslash},
\end{equation}
where $\Phi$ is a complex function, or, in the time-independent case
\begin{equation}
\psi(\vek r,E) = e^{i \chi(\vek r,E)/\hslash},
\end{equation}
where $\chi$ is also complex.
 The semiclassical approximation connects the quantum $\psi(t)$ and $\psi(E)$ to the {\emph{real}} classical action functions $S(t)$ and $W(E)$, Hamilton's principal and characteristic functions respectively. The wave functions are approximated  by the semiclassical forms, 
\begin{equation}
 \psi(\vek r,t) = e^{i \Phi(\vek r,t)/\hslash} \approx \sqrt{C(t)} \, e^{i S(\vek r,t)/\hslash},
\end{equation}
where $C(t)$ is a real function giving the density of classical trajectories.  Similarly, in the energy representation, one has
\begin{equation}
\psi(\vek r,E) = e^{i \chi(\vek r,E)/\hslash} \approx \sqrt{D(E)} \, e^{i W(\vek r,E)/\hslash},
\end{equation}
where the real function $D(E)$ expresses correspondingly the density of trajectories in energy representation. Gutzwiller, in the first two chapters of his book \cite{Gutz}, has stressed the key role in the semiclassical description played by the densities $C(t)$ and $D(E)$.

Important for scattering theory is that for free motion or motion in constant electromagnetic fields, the semiclassical approximate wave function is exact. This feature has obscured somewhat the realisation that asymptotic motion can be described semiclassically.

\subsection{Propagators in time and energy}

In time-dependent scattering theory the key element is the time propagator $U(t,t')$ transporting the system from time $t'$ to a time $t$.
This operator satisfies the TDSE
\begin{equation}
H(t)\, U(t,t') -i\hbar \frac{\partial U(t,t')}{\partial t} = 0.
\end{equation}
Here $H(t) =  H_F + V(t) = H_0  + V_F + V(t)$ where $H_0$ is the Hamiltonian (kinetic energy) operator for free motion, $V(t)$ are the possibly time-dependent interactions during the collision and $V_F$ denotes the (very weak) constant external potentials of the interaction fields. 
The operators $U_F$ and $U_0$ satisfy corresponding TDSE,
\begin{equation}
H_F\,U_F(t,t') -i\hbar \frac{\partial U_F(t,t')}{\partial t} = 0,
\end{equation}
and
\begin{equation}
H_0\,U_0(t,t') -i\hbar \frac{\partial U_0(t,t')}{\partial t} = 0.
\end{equation}
The exact quantum state propagating forward in time $(t > t')$ satisfies the time-dependent Lippman-Schwinger (LS) equation
\begin{equation}
\ket{\Psi(t)} = U_F(t,t_i)\,\ket{\psi(t_i)} - \frac{i}{\hbar}\int^t_{t_i} U(t,t')\,V(t')\ket{\psi(t')}\,dt',
\end{equation}
where $\ket{\psi(t)}$ is an eigenket of $H_F$.
Equivalently one can write
\begin{equation}
\begin{split}
\ket{\Psi(t)} &= U_F(t,t_i)\,\ket{\psi(t_i)} \\
 &- \frac{i}{\hbar}\int^t_{t_i} U_F(t,t')\,V(t')\ket{\Psi(t')}\,dt'.
\end{split}
\end{equation}

For time-independent interactions $V$, the total energy $E$ is conserved during the collision and the time propagators may be Fourier transformed to energy space. The corresponding propagators in energy space are called usually the Green operators and are defined, with forward time propagation corresponding to outgoing waves
\begin{equation}
\label{defG}
G(E) = (E - H +i\delta)^{-1},
\end{equation}
\begin{equation}
\label{defGF}
G_F(E) = (E - H_F +i\delta)^{-1},
\end{equation}
and
\begin{equation}
\label{defG0}
G_0(E) = (E - H_0 +i\delta)^{-1},
\end{equation}
where $\delta$ is a positive infinitesimal \cite{Greentime}. Again, we define $H = H_F + V \equiv H_0 + V_F + V$ where now the $V$ operator is independent of time.
Time-independent scattering theory then involves transitions between different eigenstates  of $H$ at fixed energy $E$ defined by the LS equations
\begin{equation}
\label{LSGF}
\ket{\Psi(E)} = \ket{\psi_F(E)} + G_F\,V\ket{\Psi(E)},
\end{equation}
where $\ket{\psi_F}$ is an eigenstate of $H_F$, or the alternative form
\begin{equation}
\label{LSG}
\ket{\Psi(E)} = \ket{\psi_F(E)} + G\,V\ket{\psi_F(E)}.
\end{equation}

\section{Many particles in the final channel}

\subsection{Time-dependent case}

As the IT is based on Kemble's original proof that particle momentum at the detector can be calculated from a classical velocity based upon a time of flight from microscopic reaction volume to macroscopic detector, it is appropriate to consider how this is achieved in the case of detection of several emitted particles at different arrival times.

Consider that outside the reaction volume the particles do not react with each other, although each may move in external applied fields. 
Again we take as $t=\tau$ the moment the quantum collision has ended and all reaction fragments have just exited the microscopic reaction volume. 
 
The time propagator is defined by
\begin{equation}
\ket{\Psi(t)} = U_F(t,\tau)\ket{\Psi(\tau)}
\end{equation}
with 
\begin{equation}
\label{Uonet}
U_F(t,\tau) = \exp{\left(-\frac{i}{\hslash}H_F T\right)} = \exp{\left(-\frac{i}{\hslash}\sum_n H_{Fn} \,T \right)},
\end{equation}
where $H_F$ is the total Hamiltonian of $N$ particles in the external fields and where $H_{Fn}$ is the Hamiltonian of particle $n$. Again we define the time of flight $T = t - \tau$.

The time propagator satisfies the time-dependent Schr\"odinger equation
\begin{equation}
\label{Timepropa}
\left(H_F -  i\hslash\frac{\partial}{\partial t}\right) U_F(t,\tau) = 0.
\end{equation}

The retarded energy dependent Green operator is given by
\begin{equation}
\begin{split}
\label{manyGreen}
G_F(E) &= \frac{1}{i\hslash}\int_0^\infty e^{i(E+i\delta)T/\hslash}U_F(t,\tau)~dT\\& = (E - H_F +i\delta)^{-1}\\& = \left(E - \sum_n H_{Fn} + i\delta\right)^{-1}.
\end{split}
\end{equation}
This corresponds to propagation of the $N$ particles at fixed {\emph{total}} energy $E$, the individual particle energies are not specified. Similarly the time integral implies a single time variable $T = t - \tau$ conjugate to total energy $E$.

This standard theory, however, does not correspond to the measurement of different particles at different times. 
 To reflect the conservation of individual particle energies in the asymptotic zone we propose a modification of standard theory designed to give the IT in the required form. 

The derivation rests simply on the fact that the particles move independently after leaving the collision interaction volume. 
 Then it is clear that the time propagator becomes simply a product of
single-particle propagators. However, the initial state at the edge of the reaction volume at time $\tau$ is entangled both in position and momentum space,  i.e.\ cannot be written in separable product form. 
The particles are detected at different times  $t_n$ and therefore we define a new propagator, cf.\ \eref{Uonet},
\begin{equation}
U_F(t_1,\tau,t_2,\tau,\dots)   =  \prod_n\exp{\left(-\frac{i}{\hslash}H_{Fn} T_n\right)}
\end{equation}
with $T_n = t_n - \tau$.
The state vector propagates as
\begin{equation}
\label{multit}
\begin{split}
\ket{\Psi(t_1,t_2\dots)}& = U_F(t_1,\tau,t_2,\tau, \dots) \ket{\Psi(\tau,\tau,\dots}\\
&= U_{F1}(t_1,\tau)U_{F2}(t_2,\tau) \dots \ket{\Psi(\tau,\tau,\dots)},
\end{split}
\end{equation}
where
\begin{equation}
U_{Fn}(t_n,\tau) = \exp{\left(-\frac{i}{\hslash}H_{Fn} T_n\right)}.
\end{equation}

One sees readily that this state satisfies the multiple-time Schr\"odinger equation
\begin{equation}
\begin{split}
&\sum_n \left(H_{Fn} - i\hslash\frac{\partial}{\partial t_n}\right) \ket{\Psi(t_1,t_2, \dots)}\\
& = \left(H_F - i\hslash\sum_n \frac{\partial}{\partial t_n}\right) \ket{\Psi(t_1,t_2, \dots)} = 0.
\end{split}
\end{equation}
The derivation of such multiple-time Schr\"odinger equations is discussed in \cite{JBThai}.

The analysis of section II can now be carried through for each particle separately to give the IT in the form
\begin{equation}
\label{AgainIT}
 | \Psi(\vek r_{1}(t_1), \vek r_{2}(t_2),\dots)|^2 \approx  \frac{d \vek p'}{d\vek r} \, | \tilde\Psi(\vek p'_{1},\vek p'_{2},\dots, \tau)|^2,
\end{equation}
where now
\begin{equation}
 \label{VVn}
 \frac{d \vek p'}{d\vek r}  =  \prod_{n=1}^{N}  \frac{d \vek p_{n}}{d\vek r_{n}} 
 \end{equation}
with $\vek r \equiv (\vek r_1,\vek r_2,\dots, \vek r_N)$ and $\vek p' \equiv (\vek p'_1,\vek p'_2,\dots, \vek p'_N)$.
One notes that now the ``classical" variables of the wave function propagate independently in time for each particle.
 
 For the special case of free propagation
\begin{equation}
 \label{VVnFree}
 \frac{d \vek p'}{d\vek r}  =   \prod_n  \left ( \frac{m_n}{T_n} \right)^3.
 \end{equation}

\subsection{Time-independent case}

If the time-dependence appears only in the form of energy-dependent phase factors, as it should for independent-particle propagation under time-independent Hamiltonians, 
 then
\begin{equation}
\label{factor time}
\ket{\Psi(t_1,t_2\dots)} = \ket{\Psi(E_1,E_2\dots)} \exp{\left(-\frac{i}{\hslash}\sum _n E_n T_n \right)}
\end{equation}
 with $E_n$ the energy of particle $n$. The time-independent wave function satisfies the TISE
\begin{equation}
\sum_n(H_n - E_n) \ket{\Psi(E_1,E_2\dots)}  = 0,
\end{equation}
or
\begin{equation}
\label{TISE1}
(H - E)\ket{\Psi(E)} = 0,
\end{equation}
where $E = \sum_n E_n$ is the total energy. This leads to the time-dependent form of the many-particle IT considered next.

The time-independent case involves the many-particle Green operator of \eref{manyGreen}. This is more complicated to handle as, unlike its time-dependent counterpart, it does not factorise. Only the total energy $E$ is fixed. 

Our development follows the derivation of time-independent many-particle scattering theory presented in \rref{BriggsFeaginGer}. 
For simplicity, and to connect with \rref{BriggsFeaginGer}, we consider here free-particle propagation outside the reaction zone. Otherwise, we
 will see that all quantities appearing are simple generalisations of those of Section IV.D for the three-dimensional $N=1$ case.
For $N$ particles we introduce the relative vector $\vek R \equiv \vek r - \vek r'$ with elements $\vek R_n \equiv \vek r_n - \vek r'_n$ for each particle $n$ of mass $m_n$. In \cite{BriggsFeaginGer} it was shown that, although the theory is time independent since the asymptotic motion is semiclassical, a time variable arises naturally from {\emph{time-independent quantities}}. This time is defined \cite{ephemeris}
\begin{equation}
 \label{t_W}
T =  \sqrt{ \frac{\sum_n m_n R_n^2}{2  {\cal K} }  }, 
\end{equation}
where ${\cal K}$   is the total kinetic energy of all fragments in the final channel. For simplicity of notation, in the subsequent analysis we suppress any dependence on fragment internal energies and put $ {\cal K}  \equiv E$.

The multidimensional coordinates are handled most simply by transformation to mass-weighted hyperspherical coordinates, with $3N$-dimensional vector $\vek{\mathcal R}$ and conjugate hyper-momentum
 $\vek{\mathcal P}$ (see \cite{BriggsFeaginGer} for details). The hyperradius is defined
\begin{equation}
\mathcal{R}^2 = \frac{\sum_nm_nr_n^2}{\mu},
\end{equation}
where $\mu$ is an arbitrary mass-scaling factor. Then one has
\begin{equation}
\sum_nm_nR_n^2 = \mu |\vek{\mathcal R} - \vek{\mathcal R'}|^2.
\end{equation}
Similarly the hyper-momentum at the edge of the reaction zone is related to the total energy by $\mathcal{P'} = \sqrt{2\mu E}$.

Now the characteristic action function appearing in the semiclassical Green function has the form
\begin{equation}
W(\vek r,\vek r',E) = \vek p' \cdot \vek R = \mathcal{P'}|\vek{\mathcal R} - \vek{\mathcal R}'|.
\end{equation}
With this definition it is readily shown that the time defined in \eref{t_W} corresponds to $T = \partial W/\partial E$ as should be.

Since $d\vek p_n' = (m_n/T)^3 d\vek r_n$, the Van Vleck factor \eref{VVdet} is now given by
\begin{equation}
 \label{VV_N}
\frac{d\vek p'}{d\vek r}  = \eta \left (\frac{\mu}{T} \right )^{3N}  
	= \eta \left (\frac{\mathcal P'}{|\vek{ \mathcal R} - \vek{ \mathcal R}' |} \right )^{3N} 
	=  \eta \, \frac{d\vek{ \mathcal P}'}{d\vek{ \mathcal R}},
\end{equation}
where $\vek{ \mathcal P}'$ is the initial hyper-momentum with elements $\vek{ \mathcal P}'_n = \sqrt{\mu/m_n} \, \vek p_n'$ and magnitude $\mathcal P' = 2\mu E$ and where $\eta \equiv \prod_n (m_n/\mu)^3$.
(Note $d\vek{ \mathcal R} = \eta^{1/2} \, d\vek r$ while $d\vek{ \mathcal P}' = \eta^{-1/2} \, d\vek p'$.)

Also one finds that
\begin{equation}
 \label{dtdE}
\frac{\partial^2 W}{\partial E^2} = \frac{\partial T}{\partial E} = -\mu^2 \frac{ {|\vek{ \mathcal R} - \vek{ \mathcal R}' |}}{{\mathcal P'}^{3}},
\end{equation}
so that the classical trajectory density \eref{D} is now
\begin{equation}
 \label{D_N}
\begin{split}
D = -\frac{\partial T}{\partial E} \frac{d\vek p'}{d\vek r}  &=  - \eta \, \frac{\partial T}{\partial E}  \frac{d\vek{ \mathcal P}'}{d\vek{ \mathcal R}}  \\
&= \eta \, \frac{\mu^2}{{|\vek{ \mathcal R} - \vek{ \mathcal R}' |}^2} \left (\frac{\mathcal P'}{|\vek{ \mathcal R} - \vek{ \mathcal R}' |} \right )^{3(N-1)},
\end{split}
\end{equation}
which for $N=1$ and $\mu \equiv m$ reduces to the 1-particle result $D \rightarrow m^2/R^2$.

The $N$-free-particle semiclassical Green function is then obtained as (\cite{Gutz,Berry})
\begin{equation}
 \label{G_N}
\begin{split}
&G_0(\vek{ \mathcal R}, \vek{ \mathcal R}'; E) \approx \frac{1}{i \hslash} \frac{1}{(2\pi i \hslash)^{(3N-1)/2}}  \sqrt{D} \, e^{i W/\hslash}, \\
	&= -\sqrt{2\pi} \frac{\mu}{\hslash^2}  \left (\frac{\mathcal P'}{i \hslash} \right)^{3(N-1)/2}
		\frac{ \sqrt{\eta} \,  e^{i {\mathcal P'} |\vek{ \mathcal R} - \vek{ \mathcal R}' |/\hslash}}{|\vek{ \mathcal R} - \vek{ \mathcal R}' |^{(3N-1)/2} (2\pi)^{3N/2}}.
\end{split}
\end{equation}
Again, for $N = 1$ and $\mu \equiv m$ this expression reduces to that of \eref{G0} as it should.

The $N$-particle scattering wave function \eref{exactwfnc} generalizes accordingly to 
 \begin{equation}
 \label{Psi_N}
\Psi(\vek{ \mathcal R},E)  = \int G_0(\vek{ \mathcal R}, \vek{ \mathcal R}'; E) \, V(\vek{ \mathcal R}') \, \Psi(\vek{ \mathcal R}',E) \, \eta^{-1/2} d\vek{ \mathcal R}',
\end{equation}
since $d\vek r' = \eta^{-1/2} d\vek{ \mathcal R}'$.
Introducing the asymptotic $\mathcal R \gg \mathcal R'$ limit of \eref{G_N},
\begin{equation}
 \label{G_N_asymp}
\begin{split}
G_0(\vek{ \mathcal R}, \vek{ \mathcal R}'; E) &\approx  -\sqrt{\frac{2\pi}{\hslash}} \mu \, (i \mathcal P')^{3(N-1)/2} \\
		&\times \frac{\sqrt{\eta} \,  e^{i {\mathcal P'}  {\mathcal R}/\hslash} }{R^{(3N-1)/2} } 
		\frac{ e^{-i \vek{ \mathcal P}' \cdot \vek{ \mathcal R}'/\hslash} }{ (2\pi \hslash)^{3N/2}}
\end{split}
\end{equation}
with $\vek  {\mathcal P}' \equiv {\mathcal P'} \hat{\vek {\mathcal R}}$, one obtains the asymptotic limit and $N$-particle generalization of \eref{asymp} as
\begin{equation}
 \label{Psi_N_asymp}
\Psi(\vek{ \mathcal R},E) \approx f(\vek  {\mathcal P}') \frac{e^{i {\mathcal P'}  {\mathcal R}/\hslash} }{R^{(3N-1)/2} } 
\end{equation}
with the $N$-particle scattering amplitude defined by 
\begin{equation}
 \label{f_N}
f(\vek  {\mathcal P}') \equiv  -\sqrt{\frac{2\pi}{\hslash}} \mu \, (i \mathcal P')^{3(N-1)/2} 
	\bra{\vek  {\mathcal P}' }  V \ket{\Psi}.
\end{equation}
This result defines an $N$-particle version of the differential scattering probability \eref{detectionP} and the corresponding cross section follows from a generalization of the derivation of \eref{xsec2}. See Sec.\  III.b in \cite{BriggsFeaginGer}.

In terms of the classical density of trajectories \eref{D_N}, one therefore has that
\begin{equation}
 \label{Psisq_N_D}
|\Psi(\vek{ \mathcal R},E)|^2 \approx \eta^{-1} (\mathcal P')^{-3(N-1)} \mu^{-2} D \, |f(\vek  {\mathcal P}')|^2,
\end{equation}
which is the $N$-particle generalization of \eref{ITfsq}.

Likewise, one can show that the $N$-particle generalization of \eref{fsq} connecting the scattering amplitude with the momentum wave function is given by
\begin{equation}
\label{fsqN}
|f(\vek  {\mathcal P}')|^2 \approx  (\mathcal P')^{3(N-1)} \mu^2\,  \left| \frac{\partial T}{\partial E} \right|^{-1} \,  
 |\tilde\Psi(\vek  {\mathcal P}', E)|^2, 
\end{equation}
which when combined with \eref{Psisq_N_D} gives the $N$-particle generalization of the time-independent IT, 
\begin{equation}
 \label{Psisq_N_IT}
|\Psi(\vek{ \mathcal R},E)|^2 \approx \frac{d\vek{ \mathcal P}'}{d\vek{ \mathcal R}}  |\tilde\Psi(\vek  {\mathcal P}', E)|^2. 
\end{equation}

One transforms to the laboratory coordinates $\vek r, \vek p'$ with \eref{VV_N} and noting that the momentum wave functions between laboratory and hyperspherical coordinates transform according to
$\tilde \Psi(\vek{\mathcal{P}}') =  \sqrt{\eta} \,  \tilde \Psi(\vek p')$ since $d\vek{\mathcal{ R}}' = \sqrt{\eta} \, d\vek r'$ in the Fourier transform integral. Thus,
\begin{equation}
 \label{PsisqLab_N_IT}
|\Psi(\vek r,E)|^2 \approx \frac{d\vek p'}{d\vek r}  |\tilde\Psi(\vek  p', E)|^2. 
\end{equation}

We should note in passing that the semiclassical Green function \eref{G_N} is only asymptotically equivalent for $ {\mathcal R} \gg  {\mathcal R}'$ to the quantum Green function, cf. Eq.(14) of \rref{BriggsFeaginGer}.
The reason is that the SPA derivation of the Gutzwiller expression \eref{G_N} for the $N$-particle semiclassical Green function breaks down for small $|\vek{ \mathcal R} -\vek{ \mathcal R}' |$ since there the classical density $D$ becomes rapidly varying whereas in the SPA it is assumed to be constant. (The breakdown only occurs for more than one particle.)
However, Berry and Mount \cite{Berry} have remedied this shortcoming by devising an alternate expression for the semiclassical Green function given by
\begin{equation}
 \label{G_N2}
\begin{split}
G(\vek r, \vek r'; E) &\approx -\frac{1}{i \hslash} \frac{1}{(2\pi \hslash)^{(3N-1)/2}}  \sqrt{\frac{\pi D \, W}{2\hslash}} \, H_{\alpha}^{(1)}(W/\hslash), \\
	&= -i \frac{\mu}{2\hslash^2}  \left (\frac{\mathcal P'}{2\pi \hslash} \right)^{\alpha} 
		\frac{ H_{\alpha}^{(1)}({\mathcal P'} |\vek{ \mathcal R} - \vek{ \mathcal R}' |/\hslash)}{|\vek{ \mathcal R} -\vek{ \mathcal R}' |^{\alpha} }  \sqrt{\eta},
\end{split}
\end{equation}
where $\alpha = (3N-2)/2$ and $H_{\alpha}^{(1)}$ is a Hankel function. This expression reduces to \eref{G_N} in the $ {\mathcal R} \gg  {\mathcal R}'$ limit and, given that the factor $ \sqrt{\eta}$ connects volume elements according to $d\vek{ \mathcal R}' = \sqrt{\eta} \, d\vek r'$, is precisely the {\emph{exact}} quantum Green function for free motion \cite{BriggsFeaginGer, Friedrich}.



\begin{thebibliography}{}
\bibitem{Greentime}  Note that the time propagator we designate as $U(t,t')$ is denoted as the time Green operator $G(t,t')$ by many authors.
\bibitem{TI_Maslov} As in \eref{Kpcl}, we have again suppressed for simplicity a sum over possible multiple trajectories connecting $\vek r'$ with $\vek r$ for fixed energy as well as another Maslov index that counts the number of conjugate points due to caustics along the fixed energy surface and the turning points where $V_F = E$ \cite{Gutz, Berry}.
\end{thebibliography}
\end{document}